\documentclass[letterpaper, 10 pt, conference]{ieeeconf} 
\IEEEoverridecommandlockouts  

\usepackage{times}

 \usepackage{amsmath}
 \usepackage{dsfont}
 \usepackage{graphicx}
\usepackage{multirow}
\usepackage{array} 
 \usepackage{url}
\usepackage{graphicx}
\usepackage{tikz}
\usepackage{tikz-qtree}
\usepackage[]{algorithm2e}
\usepackage{algpseudocode} 
\usepackage{amsfonts}

\newtheorem{example}{Example}
\newtheorem{remark}{Remark}
\newtheorem{problem}{Problem}

\title{\LARGE \bf  Robotic Swarm Control from Spatio-Temporal Specifications}

\author{Iman Haghighi, Sadra Sadraddini,  and Calin Belta
\thanks{This work was partially supported by the National Science Foundation under grants NRI-1426907 and CMMI-1400167.}
\thanks{I. Haghighi is with the Division of Systems Engineering, Boston University,
        Boston MA, USA
        {\tt\small haghighi@bu.edu}}%
\thanks{S. Sadraddini is with the Department of Mechanical Engineering, Boston University,
        Boston MA, USA
        {\tt\small sadra@bu.edu}}%
\thanks{C. Belta is with the Department of Mechanical Engineering,
        Boston University, Boston MA, USA
        {\tt\small cbelta@bu.edu}}%
}

\begin{document} 

\maketitle

\begin{abstract}
In this paper, we study the problem of controlling a two-dimensional robotic swarm with the purpose of achieving high level and complex spatio-temporal patterns. We use a rich spatio-temporal logic that is capable of describing a wide range of time varying and complex spatial configurations, 
and develop a method to encode such formal specifications as a set of mixed integer linear constraints, which are incorporated into a mixed integer linear programming problem. We plan trajectories for each individual robot such that the whole swarm satisfies the spatio-temporal requirements, while optimizing total robot movement and/or a metric that shows how strongly the swarm trajectory resembles given spatio-temporal behaviors. An illustrative case study is included. 
\end{abstract}

 
\section{Introduction}
\label{sec:intro}

Robotic swarms have received a lot of attention from the robotics research community in recent years \cite{mesbahi2010graph}. Large teams of robots are suitable for a broad range of applications such as distributed task allocation \cite{michael2008distributed}, area patrolling and coverage \cite{cortes2002coverage,bullo2009distributed}, search and rescue missions \cite{kantor2006distributed}, and simultaneous localization and mapping (SLAM) \cite{thrun2005multi}. With the recent technological developments, producing a large number of inexpensive robots that are equipped with sophisticated sensing, computation and communication tools has become a reality.

Describing complex spatial specifications for swarms is a non-trivial task. The existing methods rely on spatial configurations generated from simple geometrical shapes, potential fields or sets of target points \cite{chen2005formation,Pimenta:2008aa, egerstedt2001formation,lee2014multi}. However, it is practically easier to specify collective spatial behaviors of a swarm as opposed to specifying trajectories for each individual robot. The authors in \cite{yang2008multi,kloetzer2007temporal} introduced a method for controlling the abstract behavior of swarms based on the first and second moments of their spatial distribution. This is a useful approach to specify some simple patterns such as ellipsoids and boxes. However, there is a necessity for a more powerful framework of pattern specification that is not only easily definable and interpretable by the user, but is also rich enough to capture a wide range of complex spatial patterns that are not expressible by merely statistical moments. For this reason, we propose to use a formal spatial logic \cite{gol2014formal} that is capable of describing high level global behaviors in multi agent systems. 
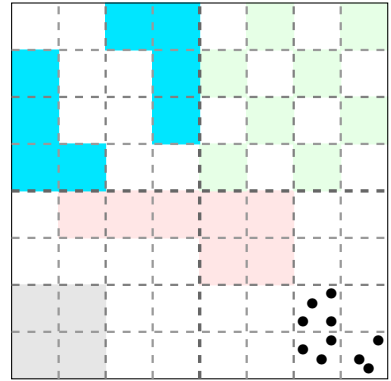
\begin{figure}[t]
\centering
\begin{tikzpicture}[xscale=1.25,yscale=1.25]

\draw[-, fill=red!10, red!10]  (0,-1) -- (0,0) -- (1,0) -- (1,-1) -- cycle;
\draw[-, fill=red!10, red!10]  (0,-0) -- (-1.5,0) -- (-1.5,-0.5) -- (0,-0.5) -- cycle;

\draw[-, fill=green!10, green!10]  (0,0) -- (0,0.5) -- (0.5,0.5) -- (0.5,0) -- cycle;
\draw[-, fill=green!10, green!10]  (0.5,0.5) -- (0.5,1) -- (1,1) -- (1,0.5) -- cycle;
\draw[-, fill=green!10, green!10]  (1.5,1.5) -- (1.5,2) -- (2,2) -- (2,1.5) -- cycle;
\draw[-, fill=green!10, green!10]  (0,1) -- (0,1.5) -- (0.5,1.5) -- (0.5,1) -- cycle;
\draw[-, fill=green!10, green!10]  (1,0) -- (1,0.5) -- (1.5,0.5) -- (1.5,0) -- cycle;
\draw[-, fill=green!10, green!10]  (1.5,0.5) -- (1.5,1) -- (2,1) -- (2,0.5) -- cycle;
\draw[-, fill=green!10, green!10]  (1,1) -- (1,1.5) -- (1.5,1.5) -- (1.5,1) -- cycle;
\draw[-, fill=green!10, green!10]  (0.5,1.5) -- (0.5,2) -- (1,2) -- (1,1.5) -- cycle;

\draw[-, fill=blue!10!cyan, blue!10!cyan]  (0,0.5) -- (0,2) -- (-0.5,2) -- (-0.5,0.5) -- cycle;
\draw[-, fill=blue!10!cyan, blue!10!cyan]  (-0.5,1.5) -- (-0.5,2) -- (-1,2) -- (-1,1.5) -- cycle;
\draw[-, fill=blue!10!cyan, blue!10!cyan]  (-1,0.5) -- (-1,0) -- (-1.5,0) -- (-1.5,0.5) -- cycle;
\draw[-, fill=blue!10!cyan, blue!10!cyan]  (-1.5,0) -- (-1.5,1.5) -- (-2,1.5) -- (-2,0) -- cycle;

\draw[-, fill=black!10!white, black!10!white]  (-1,-1) -- (-1,-2) -- (-2,-2) -- (-2,-1) -- cycle;

\draw[-]  (-2,-2) -- (-2,2) -- (2,2) -- (2,-2) -- cycle;

\draw[line width=0.3mm, dashed,color=black!50]  ( -1 ,-2) -- (-1 ,2);
\draw[line width=0.45mm, dashed,color=black!60]  ( 0 ,-2) -- (0 ,2);
\draw[line width=0.3mm, dashed,color=black!50]  ( 1 ,-2) -- (1 ,2);
\draw[line width=0.3mm, dashed,color=black!50]  ( -2 ,-1) -- (2 ,-1);
\draw[line width=0.45mm, dashed,color=black!60]  ( -2 ,0) -- (2 ,0);
\draw[line width=0.3mm, dashed,color=black!50]  ( -2 ,1) -- (2 ,1);

\draw[line width=0.3mm, dashed,color=black!40]  ( -0.5 ,-2) -- (-0.5 ,2);
\draw[line width=0.3mm, dashed,color=black!40]  ( -1.5 ,-2) -- (-1.5 ,2);
\draw[line width=0.3mm, dashed,color=black!40]  ( 0.5 ,-2) -- (0.5 ,2);
\draw[line width=0.3mm, dashed,color=black!40]  ( 1.5 ,-2) -- (1.5 ,2);

\draw[line width=0.3mm, dashed,color=black!40]  ( -2 ,-1.5) -- (2 ,-1.5);
\draw[line width=0.3mm, dashed,color=black!40]  ( -2 ,-0.5) -- (2 ,-0.5);
\draw[line width=0.3mm, dashed,color=black!40]  ( -2 ,0.5) -- (2 ,0.5);
\draw[line width=0.3mm, dashed,color=black!40]  ( -2 ,1.5) -- (2 ,1.5);

\draw (1.2,-1.2) node[] {\textcolor{black} \textbullet};
\draw (1.1,-1.4) node[] {\textcolor{black} \textbullet};
\draw (1.4,-1.1) node[] {\textcolor{black} \textbullet};
\draw (1.7,-1.8) node[] {\textcolor{black} \textbullet};
\draw (1.9,-1.6) node[] {\textcolor{black} \textbullet};
\draw (1.3,-1.8) node[] {\textcolor{black} \textbullet};
\draw (1.4,-1.6) node[] {\textcolor{black} \textbullet};
\draw (1.1,-1.7) node[] {\textcolor{black} \textbullet};
\draw (1.4,-1.4) node[] {\textcolor{black} \textbullet};
\draw (1.8,-1.9) node[] {\textcolor{black} \textbullet};

\end{tikzpicture}
\caption{An example of spatio-temporal patterning requirements for a swarm: While avoiding the unsafe zone (red) at all times, attain the following formations in any order within 30 seconds: 1)  form a checkerboard pattern (green) by  populating every other cell on the north east quadrant of the workspace. 2) Populate one of the grey squares in the south west quadrant. After completing both tasks,  gather in one of the L shaped upload regions (cyan). All the tasks must be completed within 40 seconds and each formation must be maintained for at least 3 seconds. }
\label{fig:intro}
\end{figure}

In this paper, we consider square-shaped two dimensional workspaces that are gridded to equal-sized cells as illustrated in Fig. \ref{fig:intro}. A user can express spatial requirements for the swarm by defining shapes that can be formed by unions of cells in the grid. The user can give the swarm choices between distinct patterns. Furthermore, they can also specify certain requirements for how these patterns should evolve over time. An example of such a spatio-temporal specification is given in Fig. \ref{fig:intro}. Such specifications involve logical reasoning and provide different choices for the swarm movement.  A wide variety of complex patterns can be defined in this framework that are not easily expressible by earlier work in the literature.  However, specifications of this type can be naturally expressed as spatial temporal logic (SpaTeL) \cite{haghighi2015spatel} formulas.

  

We formulate the swarm motion planning problem corresponding to a SpaTeL specification as a mixed integer linear programming (MILP) problem. A feasible solution to this problem provides a high-level plan for the movements of the swarm. We are also able to optimize a cost function. For instance, we are able to minimize the total swarm movement (energy), or find a plan that maximizes 
the satisfaction of a specification (robustness), or a combination of both. Finally, we develop a low-level strategy to move each individual robot according to the high level plan. Two different solutions for the specification given in Fig. \ref{fig:intro} corresponding to two different initial conditions are given in Fig. \ref{fig:case1} and \ref{fig:case2}. It can be seen that the optimal solution in Fig. \ref{fig:case1} involves populating the grey region before forming the checkerboard pattern and populating the right cyan region at the end, while the optimal solution in Fig. \ref{fig:case2} forms the checkerboard pattern first and populates the left cyan region.

Although temporal logic specifications in the context of mobile robot control and motion planning have been recently explored in the literature, there is very limited prior work in which complex requirements are expressed in \emph{both} space and time. The authors in \cite{kloetzer2007temporal} attempt to solve a similar problem, but spatial specifications are limited to statistical moments of the swarm in their work and thus complex spatial patterns are not easily expressible. The authors in \cite{winfield2005formal} introduced a procedure to specify emergent spatial behaviors in swarms by linear temporal logic and used model checking techniques to verify such behaviors in swarms, but the control problem is not discussed in that work. 
As opposed to linear temporal logic multi-robot motion planning \cite{chen2012formal,tumova2014receding,ulusoy2012robust,diaz2015correct}, our solution is optimization-based which is advantageous in the following ways. First, we are able to optimize a general cost function, which is difficult to formalize in automata-based approaches. We are also able to deal with infeasibility by minimizing the \emph{distance} of the swarm trajectory from satisfaction. Furthermore, under some relaxations, the complexity of our approach is independent of the size of the swarm. Therefore, our approach is easily applicable to large swarms.

This paper is organized as follows. First, we provide the necessary background on SpaTeL specifications in Sec. \ref{sec:preliminaries}. Next, the problem is formulated in Sec. \ref{problemstatement}. The solution and the technical details are explained in Sec. \ref{solution}. Finally, an illustrative case study is presented in Sec. \ref{sec:case}.

\section{Preliminaries}
\label{sec:preliminaries}

\subsection{Quad Transition Systems} 
\label{sec:quad}

A \emph{quad transition system} (QTS) \cite{gol2014formal,haghighi2015spatel} is a tree data structure defined as the tuple $Q(t)=(\mathcal{V},\mathcal{E},v_\iota,V_f,\mu,\mathcal{L}, l)$, where:
 $\mathcal{V}$ is the set of nodes (vertices). 
 $\mathcal{E} \subset \mathcal{V} \times \mathcal{V}$ is the set of directed edges (transitions). We say that $v_2$ is a \emph{child} of $v_1$ if and only if $(v_1,v_2) \in \mathcal{E}$; 
 $v_\iota$ is the root (the only node which is not a child of another node);
 $V_f$ is the set of leaves (nodes without children);
 $\mu: \mathcal{V} \times \mathbb{R}_{\geq 0}\rightarrow \mathbb{R}^+$ is the valuation function, designating each node a real value at any given time $t\geq 0$;
 $\mathcal{L}$ is a finite set of labels; 
 $l: \mathcal{E}\rightarrow \mathcal{L}$ is the labeling function that maps each edge to a label.  

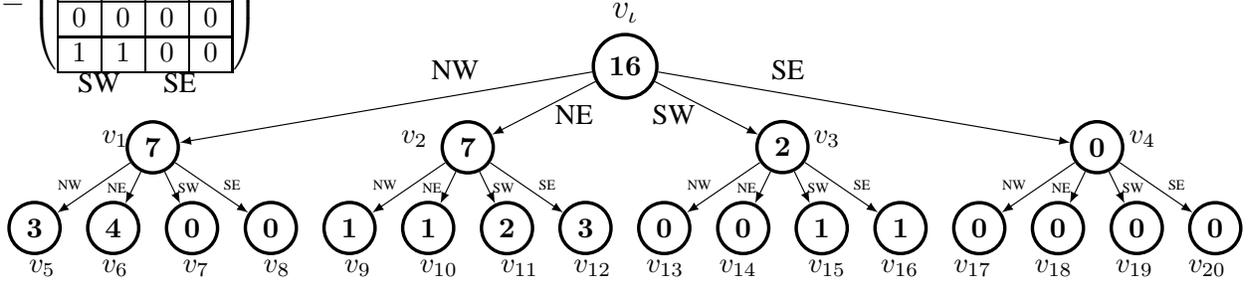
\begin{figure*}[t]
\resizebox{2\columnwidth}{!}{
\begin{tikzpicture}
[every tree node/.style={draw,circle, very thick},edge from parent/.style={draw,-latex},
   level distance=1cm,sibling distance=0.3cm, 
   edge from parent path={(\tikzparentnode) -- (\tikzchildnode)}]
\Tree [.\node[label={\large $v_\iota$}] {$\boldsymbol{16}$};
 \edge node[auto=right,near start] {NW};  
    [.$\boldsymbol{7}$
      \edge node[auto=right,near end,xshift=0.2cm] {\tiny{NW}};  
      [.$\boldsymbol{3}$  ] 
       \edge node[xshift=-0.2cm] {\tiny{NE}};
      [.$\boldsymbol{4}$ ]
      \edge node[xshift=0.2cm] {\tiny{SW}};
       [.$\boldsymbol{0}$ ]
        \edge node[auto=left, near end,xshift=-0.2cm] {\tiny{SE}};
       [.$\boldsymbol{0}$ ]
           ]
            \edge node[auto=left, near start,xshift=-0.3cm] {NE};  
    [.$\boldsymbol{7}$  
      \edge node[auto=right, near end,xshift=0.2cm] {\tiny{NW}};  
      [.$\boldsymbol{1}$  ] 
       \edge node[xshift=-0.2cm] {\tiny{NE}};
      [.$\boldsymbol{1}$ ]
      \edge node[xshift=0.2cm] {\tiny{SW}};
       [.$\boldsymbol{2}$ ]
        \edge node[auto=left,near end,xshift=-0.2cm] {\tiny{SE}};
       [.$\boldsymbol{3}$ ]
           ]
            \edge node[auto=right, near start, xshift=0.3cm] {SW};  
    [.$\boldsymbol{2}$  
      \edge node[auto=right, near end,xshift=0.2cm] {\tiny{NW}};  
      [.$\boldsymbol{0}$  ] 
       \edge node[xshift=-0.2cm] {\tiny{NE}};
      [.$\boldsymbol{0}$ ]
      \edge node[xshift=0.2cm] {\tiny{SW}};
       [.$\boldsymbol{1}$ ]
        \edge node[auto=left, near end,xshift=-0.2cm] {\tiny{SE}};
       [.$\boldsymbol{1}$ ]
           ]
            \edge node[auto=left,near start] {SE};  
    [.$\boldsymbol{0}$  
      \edge node[auto=right, near end,xshift=0.2cm] {\tiny{NW}};  
      [.$\boldsymbol{0}$  ] 
       \edge node[xshift=-0.2cm] {\tiny{NE}};
      [.$\boldsymbol{0}$ ]
      \edge node[xshift=0.2cm] {\tiny{SW}};
       [.$\boldsymbol{0}$ ]
        \edge node[auto=left, near end,xshift=-0.2cm] {\tiny{SE}};
       [.$\boldsymbol{0}$ ]
           ]
    ]
    \draw (-6.3,0.8) node[] { $A=\left( \begin{array}{|c|c|c|c|}\hline 3 & 4 & 1 & 1 \\ \hline 0 & 0 & 2 & 3 \\ \hline 0 & 0 & 0 & 0 \\ \hline 1 & 1 & 0 & 0 \\ \hline \end{array}\right)$};
    \draw (-6.5,1.8) node[] { NW};
     \draw (-5.5,1.8) node[] { NE};
     \draw (-6.5,-0.2) node[] { SW};
     \draw (-5.5,-0.2) node[] { SE};

    \draw (-6.3,-0.9) node[] { $v_1$};
    \draw (-2.6,-0.9) node[] { $v_2$};
     \draw (2.5,-0.9) node[] { $v_3$};
     \draw (6.4,-0.9) node[] { $v_4$};
       \draw (-7.2,-2.5) node[] { $v_5$};
        \draw (-6.3,-2.5) node[] { $v_6$};
         \draw (-5.3,-2.5) node[] { $v_7$};
          \draw (-4.3,-2.5) node[] { $v_8$};
           \draw (-3.3,-2.5) node[] { $v_9$};
            \draw (-2.3,-2.5) node[] { $v_{10}$};
            \draw (-1.3,-2.5) node[] { $v_{11}$};
            \draw (-0.4,-2.5) node[] { $v_{12}$};
            \draw (0.5,-2.5) node[] { $v_{13}$};
            \draw (1.4,-2.5) node[] { $v_{14}$};
            \draw (2.5,-2.5) node[] { $v_{15}$};
            \draw (3.4,-2.5) node[] { $v_{16}$};
            \draw (4.3,-2.5) node[] { $v_{17}$};
            \draw (5.3,-2.5) node[] { $v_{18}$};
            \draw (6.3,-2.5) node[] { $v_{19}$};
            \draw (7.2,-2.5) node[] { $v_{20}$};
\end{tikzpicture}
}
\caption{The QTS corresponding to the matrix $A$.}
\label{fig:ex_tree}
\end{figure*}

Let $A(t) \in \mathbb{R}^{2^D\times 2^D}$ represent a time-varying $2^D\times 2^D$ matrix, where $D \in \mathbb{N}$ is the \emph{depth} of the matrix and $t\in \mathbb{R}_{\ge 0}$ is time. 
We construct a QTS from $A(t)$ as follows. We let the \emph{root} node $v_\iota$ represent 
$A(t)$. Next, we partition the matrix into four $2^{D-1} \times 2^{D-1}$ sub-matrices, where each sub-matrix is represented by a child of $v_\iota$. We label each edge with a \emph{directional} label from the set $\mathcal{L}=\{NW,NE,SE,SW\}$, where $NW$ represents north west, $SE$ represents south east, etc (see Fig. \ref{fig:ex_tree}). Next, we execute the same  procedure for each child until the leaf nodes are obtained, i.e. each leaf is a single element matrix.
Note that $\left | V_f \right | =2^{2D}$ and $\left | \mathcal{V} \right | = \sum\limits_{i=0}^{D} 2^{2D}$. For each leaf node $v_f \in \mathcal{V}_f$, we let $\mu(v_f,t)$ to be the value of the corresponding element in $A(t)$. The valuation function for other nodes is recursively defined as the sum of the valuations of its children, i.e. 
\begin{equation}
\mu(v,t)= \sum \limits_{(v,v_c)\in\mathcal{E}} \mu(v_c,t)\;\;\; \forall v\in \mathcal{V} \setminus V_f.
\label{eqn:valuation}
\end{equation}
An example of a QTS construction is given in Fig. \ref{fig:ex_tree}.

Given a subset of labels $\mathcal{B} \subseteq \mathcal{L}$, a \emph{labeled path} of a QTS is defined as a function that maps a vertex to a set of infinite sequences of nodes:
\begin{equation}
\begin{array}{r}
\lambda^\mathcal{B}(v_0):=  \{ (v_0,v_1,v_2\cdots,\overline{v_f}) \big |(v_i,v_{i+1})\in\mathcal{E}, v_f \in V_f, \\ l(v_i,v_{i+1}) \in \mathcal{B}, i\in\mathbb{N}_{\geq 0}  \},
\end{array}
\end{equation}  
where $\overline{v_f}$ denotes infinite repetitions of leaf node $v_f$. The $i$-th element of a labeled path $\pi\in\lambda^{\mathcal{B}}(v)$ is denoted by $\pi_i$.
For example, in Fig. \ref{fig:ex_tree}, $(v_0,v_1,\overline{v_5})$ and $(v_0,v_1,\overline{v_6})$ are both members of $\lambda^{\{NW,NE\}}(v_0)$.

A \emph{QTS signal} starting at time $t$ is defined as $Q_t=\{Q(\tau)|\tau\geq t\}$.

\subsection{Spatial Temporal Logic (SpaTeL)}
\label{sec:SpaTeL}
SpaTeL formulas are defined by nesting tree spatial superposition logic (TSSL) specifications \cite{gol2014formal} inside temporal operators of signal temporal logic (STL) \cite{maler_stl}. Formal definitions of SpaTeL syntax and semantics can be found in \cite{haghighi2015spatel}. Informally, SpaTeL formulas are STL formulas in which linear predicates over signals are replaced with spatial formulas over quad transition systems and allow for describing how spatial patterns change over time.

A TSSL formula is recursively formed by linear predicates over the valuation function \eqref{eqn:valuation}, spatial operators, and boolean operators ($\wedge$,$\vee$,$\neg$). For example, $\varphi:=\mu\sim c$, where $\sim\in\{\geq,<\}$, is a very simple TSSL formula consisting of a single predicate that indicates that the function $\mu$ of \eqref{eqn:valuation} at the initial node $v_\iota$ must have a value of larger (smaller) than threshold $c$. If the predicate is true, we write $Q^{v_\iota}\models\varphi$ (read as QTS $Q$ satisfies formula $\varphi$ at node $v_\iota$).  Spatial operators are used to define specifications at nodes located in lower tree levels. For instance, $\exists_\mathcal{B}\bigcirc \varphi$ ($\mathcal{B}\subseteq\{NW,NE,SW,SE\}$) is the spatial ``there exists next'' operator which means that the spatial formula $\varphi$ has to be satisfied for at least one of the children of the initial node with directional label $l(v_\iota,v')\in\mathcal{B}$. Furthermore, $\forall_\mathcal{B}\bigcirc \varphi$, read as ``for all next'', indicates that $\varphi$ must be satisfied by all such children. Specifications at deeper tree levels can be expressed similarly by nesting several spatial next operators. In addition to spatial next, TSSL is equipped with spatial until operators ($\exists_{\mathcal{B}} \varphi_1 \mathcal{U}_\kappa \varphi_2 , \forall_{\mathcal{B}} \varphi_1 \mathcal{U}_\kappa \varphi_2$). Formal definitions for all these operators are presented in \cite{gol2014formal}.

A SpaTeL formula is recursively formed by TSSL formulae, temporal operators, and boolean operators. Three common temporal operators are eventually ($F_I$), always ($G_I$), and until ($U_I$), where $I=[t_1,t_2)$ is a time interval. For instance, A QTS signal $Q_t\models F_I\varphi$ ( read as $Q_t$ satisfies $F_I\varphi$) if there $\exists \tau\in[t+t_1,t+t_2)$ such that $Q_\tau\models\varphi$ and $Q_t\models G_I\varphi$ if $\forall \tau\in[t+t_1,t+t_2) ~Q_\tau\models\varphi$.  
 A SpaTel formula is satisfied by a QTS signal $Q_t$ if and only if it is satisfied by the QTS at time t ($Q(t)$).

\begin{example}
\label{example:tssl}
Consider a $4\times 4$ matrix with the requirement that every other entry is zero (thus forming a checkerboard pattern). There are two different realizations for this pattern, that can be specified by the following TSSL formulas:
\begin{small}
\begin{equation*}
\begin{array}{c}
\varphi_{c1}=\forall_\mathcal{L}\bigcirc(\forall_{\{NW,SE\}}\bigcirc(\mu = 0)) \\
\varphi_{c2}=\forall_\mathcal{L}\bigcirc(\forall_{\{NE,SW\}}\bigcirc(\mu = 0)),

\end{array}
\end{equation*}
\end{small}
where $\mu$ is the valuation function in the definition of QTS. Now consider a spatio-temporal requirement that the checkerboard pattern periodically switches tiles. The following SpaTeL formula specifies this requirement:
\begin{equation}
 \Phi_c=G_{[0,t_1)}(F_{[0,t_2)}\varphi_{c1}\wedge F_{[0,t_2)}\varphi_{c2}).
 \label{eqn:spatel_CB}
 \end{equation}
\end{example}

\vspace{5pt}
SpaTeL is equipped with quantitative semantics. Quantitative valuation (robustness) $\rho(\Phi,Q_t)$ of a SpaTeL formula $\Phi$ with respect to QTS signal $Q_t$ is calculated recursively:
\begin{equation}
\begin{array}{l}
\rho (\varphi, Q_t) = \rho_s (\varphi, v_\iota), \\
\rho (\neg \Phi, Q_t)  = - \rho (\Phi, Q_t),\\
\rho (\Phi_1 \wedge \Phi_2, Q_t)= \min (\rho(\Phi_1, Q_t), \rho (\Phi_2, Q_t)) , \\
\rho (\Phi_1 \vee \Phi_2, Q_t) = \max (\rho(\Phi_1, Q_t), \rho (\Phi_2, Q_t)), \\
\rho(F_{I_1,I_2)}\Phi,Q_t)=\sup_{t'\in[t+I_1,t+I_2)}\rho(\Phi,Q_{t'}), \\
\rho(G_{I_1,I_2)}\Phi,Q_t)=\inf_{t'\in[t+I_1,t+I_2)}\rho(\Phi,Q_{t'}), \\
\rho (\Phi_1 U_{[I_1, I_2)} \Phi_2, Q_t) = \sup_{t' \in [t+I_1, t+I_2)}  ( \\ \min (\rho (\Phi_2, Q_{t'}),\inf_{t'' \in [t+I_1, t')} \rho (\Phi_1, Q_{t''}))),
\end{array}
\label{eqn:quantitative_semantics}
\end{equation}
where $\rho_s(\varphi,v)$ is the robustness of TSSL formula $\varphi$ with respect to node $v\in\mathcal{V}$:

\begin{equation}
\label{eqn:quantitative_semantics_2}
\begin{array}{l}
\rho_s (\top, v) = 1 , \\
\rho_s (\mu \sim c, v) =  (\mu-c)\text{ if }(\sim \text{ is } \geq) ,(c-\mu)\text{ if }(\sim\text{is}\leq), \\
\rho_s (\neg \varphi, v) = - \rho_s (\varphi, v),\\
\rho_s (\varphi_1 \wedge \varphi_2, v) = \min (\rho_s (\varphi_1, v), \rho_s (\varphi_2, v)), \\
\rho_s (\varphi_1 \vee \varphi_2, v) = \max (\rho_s (\varphi_1, v), \rho_s (\varphi_2, v)),\\
\rho_s (\exists_\mathcal{B} \bigcirc \varphi, v) =  \max_{\pi \in \lambda^\mathcal{B} (v)} \rho_s (\varphi,\pi_1), \\
\rho_s (\forall_\mathcal{B} \bigcirc \varphi, v) =  \min_{\pi \in \lambda^\mathcal{B} (v)} \rho_s (\varphi,\pi_1),\\
\rho_s (\exists_\mathcal{B} \varphi_1 U_k \varphi_2, v) = \sup_{\pi \in \lambda^\mathcal{B} (v), i \in (0, k]}( \\ \min (\rho_s (\varphi_2, \pi_i), \inf_{j \in [0, i)}  \rho_s (\varphi_1, \pi_j))),\\
\rho_s (\forall_\mathcal{B} \varphi_1 U_k \varphi_2, v) = \inf_{\pi \in \lambda^\mathcal{B}(v), i \in (0, k]}( \\ \min (\rho_s (\varphi_2, \pi_i), \inf_{j \in [0, i)}  \rho_s (\varphi_1, \pi_j))).
\end{array}
\end{equation}

Positive and negative robustness indicate satisfaction and violation, respectively.
\begin{equation}
\label{eqn:sound}
\begin{array}{l}
\rho(\Phi,Q_t)>0\Rightarrow Q_t \models \Phi, \\
\rho(\Phi,Q_t)<0\Rightarrow Q_t \not \models \Phi.
\end{array}
\end{equation}
  The absolute robustness value can be  
 viewed as a measure of "distance to satisfaction". In other words, a higher absolute value for robustness indicates stronger satisfaction (violation) of a specification. We use the definition of robustness (\eqref{eqn:quantitative_semantics}, \eqref{eqn:quantitative_semantics_2}) in subsequent sections to translate SpaTeL specifications into mixed integer constraints.
 
 \begin{example}[Example \ref{example:tssl} continued]
  Consider a stationary QTS $Q(t)=\mathcal{Q}$, where $\mathcal{Q}$ is the QTS depicted in Fig. \ref{fig:ex_tree}. By computing quantitative semantics from \eqref{eqn:quantitative_semantics_2}, it is straightforward to verify that specification $\Phi_c$ in \eqref{eqn:spatel_CB} is violated by $Q_0$ with a robustness of $-4$.
  \end{example}
 
The horizon of a SpaTeL formula is defined similar to STL \cite{dokhanchi}. Intuitively, the horizon $T$ of a SpaTeL formula $\Phi$ is the maximum time for which some specification in $\Phi$ must be checked against $Q_t$. For instance, the time horizon of $G_{[0,20)}F_{[0,5)}\forall_\mathcal{L}\bigcirc (\mu \geq1)$ is $T=25$. 
\section{Problem Formulation and Approach}
\label{problemstatement}

Consider $N$ homogenous planar robots with negligible sizes in a two-dimensional space. The position of robot $r$ at time $t$ is denoted by $x_r(t) \in \mathcal{X}$, $r=1,\cdots,N$, where $\mathcal{X} \subset \mathbb{R}^2$ is the workspace of the robots, which is assumed to be the following square:
$\mathcal{X} :=[-\frac{a}{2},\frac{a}{2}] \times [-\frac{a}{2},\frac{a}{2}],$
where $a$ is the length of the square. Note that any rectangular workspace can be normalized to meet this assumption. We denote the state of the swarm by $x(t)=\left(x_1(t)^T,x_2(t)^T,\cdots, x_N(t)^T\right)^T$. The kinematics of each individual robot is assumed as follows:
\begin{equation}
\label{eqn:model}
\dot{x}_r(t)=u_r(t),\;\;\; r=1,\cdots,N,
\end{equation}
where $u_r(t) \in \mathcal{U}$ is the control applied to robot $r$ at time $t$ and $\mathcal{U}= \left \{ u_r \big | \left \|u \right \|_2 \le u_m \right\}$, where $u_m$ is the maximum speed that a robot can attain. 
 
$\mathcal{X}$ is partitioned into $2^D \times 2^D$ number of equal-sized cells, where $D$ is the depth of the grid. A user expresses desirable patterns by defining shapes that are formed by unions of cells in the workspace, defining thresholds for the number of agents populating each shape, and expressing temporal requirements for those pre-identified patterns. The objective in this paper is to synthesize a control policy for \eqref{eqn:model} such that the spatio-temporal requirements expressed by the user are met. We will provide a formal formulation for this problem later in this section. 

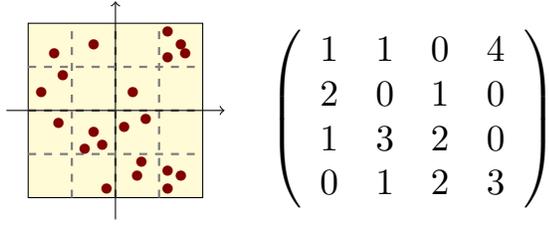
\begin{figure}
\centering
\begin{tabular}{lll}
\begin{tikzpicture}[xscale=0.58,yscale=0.58]
\draw[-, fill=yellow!20]  (-2,-2) -- (-2,2) -- (2,2) -- (2,-2) -- cycle;
\draw[->] (-2.5,0) -- (2.5,0) ;
\draw[->] (0,-2.5) -- (0,2.5);
\draw[line width=0.3mm, dashed,color=black!50]  ( -1 ,-2) -- (-1 ,2);
\draw[line width=0.3mm, dashed,color=black!80]  ( 0 ,-2) -- (0 ,2);
\draw[line width=0.3mm, dashed,color=black!50]  ( 1 ,-2) -- (1 ,2);
\draw[line width=0.3mm, dashed,color=black!50]  ( -2 ,-1) -- (2 ,-1);
\draw[line width=0.3mm, dashed,color=black!80]  ( -2 ,0) -- (2 ,0);
\draw[line width=0.3mm, dashed,color=black!50]  ( -2 ,1) -- (2 ,1);

\draw (1.5,1.5) node[] {\textcolor{red!50!black} \textbullet};
\draw (1.6,1.3) node[] {\textcolor{red!50!black}\textbullet};
\draw (-1.4,1.3) node[] {\textcolor{red!50!black} \textbullet};
\draw (0.4,0.4) node[] {\textcolor{red!50!black} \textbullet};
\draw (0.2,-0.4) node[] {\textcolor{red!50!black} \textbullet};
\draw (1.2,-1.4) node[] {\textcolor{red!50!black} \textbullet};
\draw (0.7,-0.2) node[] {\textcolor{red!50!black} \textbullet};
\draw (-0.7,-0.9) node[] {\textcolor{red!50!black} \textbullet};
\draw (-1.7,0.4) node[] {\textcolor{red!50!black} \textbullet};
\draw (-1.2,0.8) node[] {\textcolor{red!50!black} \textbullet};
\draw (-1.3,-0.3) node[] {\textcolor{red!50!black} \textbullet};
\draw (-0.3,-0.8) node[] {\textcolor{red!50!black} \textbullet};
\draw (-0.2,-1.8) node[] {\textcolor{red!50!black} \textbullet};
\draw (-0.5,-0.5) node[] {\textcolor{red!50!black} \textbullet};
\draw (0.5,-1.5) node[] {\textcolor{red!50!black} \textbullet};
\draw (-0.5,1.5) node[] {\textcolor{red!50!black} \textbullet};
\draw (1.5,-1.5) node[] {\textcolor{red!50!black} \textbullet};
\draw (1.2,-1.8) node[] {\textcolor{red!50!black} \textbullet};
\draw (0.6,-1.2) node[] {\textcolor{red!50!black} \textbullet};
\draw (1.2,1.2) node[] {\textcolor{red!50!black} \textbullet};
\draw (1.2,1.8) node[] {\textcolor{red!50!black} \textbullet};

\end{tikzpicture}
&
\begin{tikzpicture}[xscale=0.9,yscale=1.7]
%
%

\draw (0,-0.3) node[scale=1.4] {
$\left(
\begin{array}{cccc}
1 & 1 & 0 & 4 \\
2 & 0 & 1 & 0 \\
1 & 3 & 2 & 0 \\
0 & 1 & 2 & 3 \\
\end{array}
\right)
$
};
\end{tikzpicture}

\end{tabular}

\caption{(Left) A swarm of $21$ robots in a square region. The square is gridded  into 16 cells. (Right) The matrix representing the number of the robots in each cell. }
\label{fig:robots}
\end{figure}

We construct the matrix $\mathcal{N}(t) \in \mathbb{N}^{2^D \times 2^D}$, where the value of each element is the number of robots in the corresponding cell, as illustrated in an example in Fig. \ref{fig:robots}. We construct the time varying QTS $Q(t)$ from $\mathcal{N}(t)$ using the procedure outlined in Sec. \ref{sec:quad}. Note that the shapes defined by unions of cells can be easily expressed using the spatial next operator in tree spatial superposition logic (see Example \ref{ex:spec}). Consequently, a SpaTeL specification $\Phi$ can be automatically generated from the input specification.
\begin{remark}
A supervised learning algorithm is proposed in \cite{gol2014formal} for automatically learning TSSL formulae that are satisfied by a set of positive spatial configurations (images) and violated by a set of negative images. This method can be used to learn TSSL (and SpaTeL) formulas describing more complex high level patterns (circular clusters, ellipsoids, etc). Two sets of images, one illustrating the desirable pattern and the other lacking the pattern, should be artificially created. The learning algorithm generates a TSSL formula by using the generated images as a training set. Although this is a very effective method to find SpaTeL descriptors for arbitrary patterns, the resulting formulas are often too long and complex. Therefore, The mixed integer linear programming problems that result from the framework presented in subsequent sections become unsolvable by existing solvers. As a result, the input spatio-temporal requirements in this paper are limited to unions of squares. As explained in Sec. \ref{sec:SpaTeL}, such requirements can be intuitively translated into TSSL/SpaTeL specifications and the resulting formulae are small and manageable.  
\end{remark}

\begin{example}
\label{ex:spec}
The specification that was introduced in Sec. \ref{sec:intro} (Fig. \ref{fig:intro}) is formalized by the following SpaTeL formula:
\begin{equation}
\begin{array}{r}
\Phi_d=G_{[0,40)}(\neg\varphi_1)\wedge(F_{[0,30)}G_{[0,3)}\varphi_2) \\ \wedge(F_{[0,30)}G_{[0,3)}\varphi_3)\wedge(F_{[30,40)}\varphi_4),
\end{array}
\label{eqn:case_spatel}
\end{equation}
where $\varphi_i$ are TSSL formulas describing different patterns illustrated in Fig. \ref{fig:intro}: $\varphi_1$ represents the red danger zone in Fig. \ref{fig:intro}, $\varphi_2$ specifies formation of a checkerboard pattern in the north west quadrant, $\varphi_3$ specifies gathering inside one of the grey cells in the south west quadrant, and $\varphi_4$ represents populating one of the L-shaped cyan regions. These formulas are automatically generated by representing each cell in the gridded workspace using appropriate spatial next operators of TSSL. 
\begin{equation}
\begin{array}{ll}
\varphi_1= &\forall_{SE}\bigcirc\forall_{NW}\bigcirc(\mu\leq 0)\wedge \\&
\forall_{SW}\bigcirc\forall_{NE}\bigcirc\forall_{\{NW,NE\}}\bigcirc (\mu \leq 0)\wedge \\&
\forall_{SW}\bigcirc\forall_{NW}\bigcirc\forall_{NE}\bigcirc (\mu \leq 0), \\
\varphi_2= &\forall_{NE}\bigcirc(\forall_\mathcal{L}\bigcirc\forall_{\{NW,SE\}}\bigcirc(\mu\geq \gamma_1)), \\
\varphi_3= &\forall_{SW}\bigcirc(\forall_{SW}\bigcirc\exists_\mathcal{L}\bigcirc(\mu\geq \gamma_2)), \\
\varphi_4= &\forall_{NW}\bigcirc(\varphi_5\vee\varphi_6), \\
\varphi_5= &(\forall_{NE}\bigcirc\forall_{\{NW,NE,SE\}}\bigcirc(\mu\geq \gamma_3))\wedge \\ &(\forall_{SE}\bigcirc\forall_{NE}\bigcirc(\mu\geq \gamma_4)), \\
\varphi_6= &(\forall_{SW}\bigcirc\forall_{\{NW,SW,SE\}}\bigcirc(\mu\geq \gamma_5))\wedge \\ &(\forall_{NW}\bigcirc\forall_{SW}\bigcirc(\mu\geq \gamma_6)),
\end{array}
\end{equation}
where $\mathcal{L}=\{NW,NE,SW,SE\}$, and $\mu$ is the the number of robots residing in a subregion of the workspace identified by spatial operators and $\gamma_{1-6}$ are thresholds for the minimum number of robots that are required to populate each pattern.
\end{example}

 We wish to find a control strategy that steers the swarm such that $\Phi$ is satisfied. Such a policy is not usually unique. Therefore, we choose a policy that optimizes a cost function. For instance, we can minimize the total number of robot displacements (one displacement is defined as moving one robot from its current location to a neighboring cell). In addition, a natural candidate for optimization is maximizing the SpaTeL robustness. The problem that we consider in this paper is formulated as follows:

\begin{problem}
\label{problem1}
Given a swarm of $N$ agents with initial positions at $x(0)$ and a SpaTeL formula $\Phi$ that describes time varying spatial requirements of the user, find an \emph{optimal and correct} control strategy such that:
\begin{equation}
\begin{array}{ccl}
\underset{r=1,\cdots,N, t \in [0,T]} {u_r(t)}  = & argmin &  -\alpha~ \rho(\Phi,Q_0)+  J_f(x(T)) \\&& + \int_0^T J_r(x(t),u(t)) dt, \\

& s.t. & Q_0 \models \Phi,
\end{array}
\end{equation}
where $\rho$ is the SpaTeL robustness, $Q_0$ is the QTS signal starting at time $0$, $J_f: \mathbb{R}^{2N} \rightarrow \mathbb{R}$, $J_r: \mathbb{R}^{2N} \times \mathbb{U}^N \rightarrow \mathbb{R}$, are the endpoint cost and the running cost (Lagrangian), respectively. The end time $T$ is the time horizon of the SpaTeL formula and $\alpha$ is a positive constant designating a weight for SpaTeL robustness.
\end{problem}

Our approach to problem \ref{problem1} can be summarized as follows. First, we find $\mathcal{N}(t), 0\le t \le T$ such that $Q_0 \models \Phi$. It is known that this problem is undecidable in continuous time \cite{raman}. Therefore, we (approximately) solve the problem in discrete time assuming that at each time step, each robot can be displaced by one cell to its right, left, up or down. Therefore, we choose a sampling time such that:
$\Delta t \ge \frac{a}{2^{D-1}u_m}.$
We assume that the time intervals of the temporal operators of $\Phi$ are multiples of $\Delta t$. This assumption can be matched by increasing $D$ such that the time intervals can be reasonably approximated by multiples of $\Delta t$.
We also denote the last discrete time by $K:=\frac{T}{\Delta t}$.   
The matrix $\mathcal{N}$ at time $t=k\Delta t$ is denoted by $\mathcal{N}[k]$. We construct a discrete time model for the evolution of $\mathcal{N}[k]$. Next, we find the required values at each time such that the SpaTeL specification is satisfied using a MILP-based approach that is explained in Sec. \ref{solution}. Finally, we find continuous time controls for each individual robot such that the number of each cell at time $t=k \Delta t$ matches its corresponding value in $\mathcal{N}[k]$.

\section{Solution}
\label{solution}

\subsection{Swarm Flow in Discrete Time}
\begin{figure}[t]
\centering
\scriptsize
\begin{tikzpicture}[scale=1]
    \node[shape=circle,draw=black, fill=yellow!50] (A) at (0,0) {$4,1$};
    \node[shape=circle,draw=black, fill=yellow!50] (B) at (0,1) {$3,1$};
    \node[shape=circle,draw=black, fill=yellow!50] (C) at (0,2) {$2,1$};
    \node[shape=circle,draw=black, fill=yellow!50] (D) at (0,3) {$1,1$};
    \node[shape=circle,draw=black, fill=yellow!50] (E) at (1,3) {$1,2$};
    \node[shape=circle,draw=black, fill=yellow!50] (F) at (1,2) {$2,2$} ;
    \node[shape=circle,draw=black, fill=yellow!50] (G) at (1,1) {$3,2$} ;
    \node[shape=circle,draw=black, fill=yellow!50] (H) at (1,0) {$4,2$} ;
    
    \node[shape=circle,draw=black, fill=yellow!50] (A1) at (2,0) {$4,3$};
    \node[shape=circle,draw=black, fill=yellow!50] (B1) at (2,1) {$3,3$};
    \node[shape=circle,draw=black, fill=yellow!50] (C1) at (2,2) {$2,3$};
    \node[shape=circle,draw=black, fill=yellow!50] (D1) at (2,3) {$1,3$};
    \node[shape=circle,draw=black, fill=yellow!50] (E1) at (3,3) {$1,4$};
    \node[shape=circle,draw=black, fill=yellow!50] (F1) at (3,2) {$2,4$} ;
    \node[shape=circle,draw=black, fill=yellow!50] (G1) at (3,1) {$3,4$} ;
    \node[shape=circle,draw=black, fill=yellow!50] (H1) at (3,0) {$4,4$} ;

    \path [->, line width=0.4 mm,bend left](A) edge node[] {} (B);
    \path [<-, line width=0.4 mm,bend right](A) edge node[] {} (B);
    \path [->, line width=0.4 mm,bend left](B) edge node[] {} (C);
     \path [<-, line width=0.4 mm,bend left,bend right](B) edge node[] {} (C);
    \path [->, line width=0.4 mm,bend left](C) edge node[] {} (D);
    \path [<-, line width=0.4 mm,bend right](C) edge node[] {} (D);
    \path [->, line width=0.4 mm,bend left](D) edge node[] {} (E);
     \path [<-, line width=0.4 mm,bend right](D) edge node[] {} (E);
    \path [->, line width=0.4 mm,bend left](E) edge node[] {} (F);
    \path [<-, line width=0.4 mm,bend right](E) edge node[] {} (F);
    \path [->, line width=0.4 mm,bend left](F) edge node[] {} (G);
     \path [<-, line width=0.4 mm,bend right](F) edge node[] {} (G);
    \path [->, line width=0.4 mm,bend left](G) edge node[] {} (H);
     \path [<-, line width=0.4 mm,bend right](G) edge node[] {} (H);
    \path [->, line width=0.4 mm,bend left](A) edge node[] {} (H);
     \path [<-, line width=0.4 mm,bend right](A) edge node[] {} (H);
    \path [->, line width=0.4 mm,bend left](B) edge node[] {} (G);
    \path [<-, line width=0.4 mm,bend right](B) edge node[] {} (G);
    \path [->, line width=0.4 mm,bend left](C) edge node[] {} (F);
     \path [<-, line width=0.4 mm,bend right](C) edge node[] {} (F);
    
    \path [->, line width=0.4 mm,bend left](A1) edge node[] {} (B1);
    \path [<-, line width=0.4 mm,bend right](A1) edge node[] {} (B1);
    \path [->, line width=0.4 mm,bend left](B1) edge node[] {} (C1);
     \path [<-, line width=0.4 mm,bend right](B1) edge node[] {} (C1);
    \path [->, line width=0.4 mm,bend left](C1) edge node[] {} (D1);
    \path [<-, line width=0.4 mm,bend right](C1) edge node[] {} (D1);
    \path [->, line width=0.4 mm,bend left](D1) edge node[] {} (E1) edge node[] {} (E);
    \path [<-, line width=0.4 mm,bend right](D1) edge node[] {} (E1) edge node[] {} (E);
    \path [->, line width=0.4 mm,bend left](E1) edge node[] {} (F1);
    \path [<-, line width=0.4 mm,bend right](E1) edge node[] {} (F1);
    \path [->, line width=0.4 mm,bend left](F1) edge node[] {} (G1);
    \path [<-, line width=0.4 mm,bend right](F1) edge node[] {} (G1);
    \path [->, line width=0.4 mm,bend left](G1) edge node[] {} (H1);
     \path [<-, line width=0.4 mm,bend right](G1) edge node[] {} (H1);
    \path [->, line width=0.4 mm,bend left](A1) edge node[] {} (H1) edge node[] {} (H);
    \path [<-, line width=0.4 mm,bend right](A1) edge node[] {} (H1) edge node[] {} (H);
    \path [->, line width=0.4 mm,bend left](B1) edge node[] {} (G1) edge node[] {} (G);
    \path [<-, line width=0.4 mm,bend right](B1) edge node[] {} (G1) edge node[] {} (G);
    \path [->, line width=0.4 mm,bend left](C1) edge node[] {} (F1) edge node[] {} (F);
    \path [<-, line width=0.4 mm,bend right](C1) edge node[] {} (F1) edge node[] {} (F);

\end{tikzpicture}
\normalsize
\caption{A flow network with $16$ cells. The robots are only able to move from one cell to a neighboring cell in one time step.}
\label{fig:flow}
\end{figure}
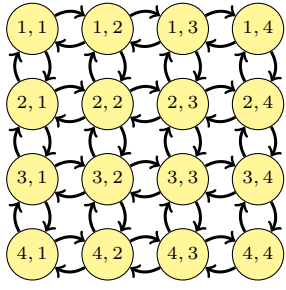
In this section, we develop a discrete time model that characterizes the evolution of $\mathcal{N}[k]$. At each time step, each robot is only able to remain at its current cell or move to an adjacent cell (The cell to its right, left, top, or down). All the robots move synchronously during one time step. The flow of the robots between the cells can be thought as a network as depicted in Fig. \ref{fig:flow}. The index of each cell is represented by $[i,j]$, where $i$ is the row and $j$ is the column of the element in the matrix $\mathcal{N}[k]$.  The set of cells that are adjacent to $[i,j]$ is denoted by $\Omega([i,j])$.
We denote the number of robots in the cell $[i,j]$ at time step $k$ by $\mathcal{N}_{[i,j]}[k]$. The number of robots that move from cell $[i,j]$ to an adjacent cell $[i',j']\in\Omega([i,j])$ during time $[k, {k+1}]\Delta t$ is denoted by $f_{[i,j]}^{[i^\prime,j^\prime]}[k]$, which is a non-negative integer. 
The total number of robots that move out from cell $[i,j]$ at time step $k$ is:
\begin{equation}
f^{out}_{[i,j]}[k]:= \underset{[i',j']\in\Omega([i,j])} {\sum} f_{[i,j]}^{[i',j']}[k].
\end{equation} 
We add the following constraint:
\begin{equation}
\mathcal{N}_{[i,j]}[k] \ge f^{out}_{[i,j]}[k],
\end{equation}
which indicates that the number of robots moving out from a cell can not be more than the number of robots in the cell. The number of robots that enter cell $[i,j]$ at $k$ is:
\begin{equation}
f^{in}_{[i,j]}[k]:= \underset{[i',j']\in \Omega([i,j])} {\sum} f_{[i',j']}^{[i,j]}[k].
\end{equation} 
The discrete time evolution of $\mathcal{N}[k]$ is:
\begin{equation}
\mathcal{N}_{[i,j]}[k+1]=\mathcal{N}_{[i,j]}[k] - f^{out}_{i,j}[k] + f^{in}_{i,j}[k],   
\end{equation}
which is a function of decisions made on the values of $f_{[i,j]}^{[i^\prime,j^\prime]}[k]$. In a compact form, we define the decision variable $f[k]$ as the set:
\begin{equation}
f[k]=\left \{  f_{[i,j]}^{[i',j']} \big| \forall [i',j']\in\Omega([i,j]),\forall [i,j]\right\},
\end{equation}
and the discrete time evolution of $\mathcal{N}[k]$ is written as:
\begin{equation}
\mathcal{N}[k+1]=\mathcal{F}(\mathcal{N}[k], f[k]).
\end{equation} 

\subsection{Mixed-Integer Formulation of SpaTeL Specifications}
\label{sec:mixed}

In this section, we explain how to recursively transform a SpaTeL formula into a set of mixed-integer constraints. Our method is inspired by the binary mixed-integer encoding of STL formulas presented in \cite{raman}.

For a predicate of a SpaTeL formula $\sigma=(\mu \geq c)$, a set of binary variables $z_\sigma[v,k] \in \{0,1\}$, $v\in\mathcal{V}$, $0\leq k \leq K$, is associated such that values 1 and 0 indicate $True$ and $False$, respectively. The corresponding mixed integer constraints are:
\begin{equation}
\label{eqn:predicate}
\left \{ 
\begin{array}{ll}
\mu[v,k]-Mz_\sigma[v,k] & \le c, \\
\mu[v,k]+M(1-z_\sigma[v,k]) & \geq c,
\end{array}
\right.
\end{equation}
where $M$ is a sufficiently large positive number. Mixed integer constraints for all predicates in the form of $\sigma'=(\mu\leq c)$ are defined similarly.
For encoding a SpaTeL formula, The following rules are used to map boolean, temporal, and spatial operators into mixed integer constraints. These rules are derived from the definition of SpaTeL robustness \eqref{eqn:quantitative_semantics},\eqref{eqn:quantitative_semantics_2}.

\begin{itemize}
\item \emph{Negation}: $$\Psi=\neg\Phi\rightarrow z_\Psi[v,k]=1-z_\Phi[v,k];$$
\item \emph{Conjunction}: $$\Psi=\bigwedge_{i=1}^m \Phi_i\rightarrow\left\{ 
\begin{array}{l}
z_\Psi[v,k] \le z_{\Phi_i}[v,k], i=1,\cdots,m, \\
z_\Psi[v,k] \ge 1-m+\sum \limits_{i=1}^m z_{\Phi_i}[v,k];
\end{array}
\right.$$
\item \emph{Disjunction}: $$\Psi=\bigvee_{i=1}^m \Phi_i\rightarrow\left\{ 
\begin{array}{l}
z_\Psi[v,k] \ge z_{\Phi_i}[v,k], \\
z_\Psi[v,k] \le \sum \limits_{i=1}^m z_{\Phi_i}[v,k];
\end{array}
\right.$$
\item \emph{There exists spatial next}: $$\Psi=\exists_{\mathcal{B}} \bigcirc \Phi\rightarrow z_\Psi[v,k] = \bigvee_{\pi \in \lambda^\mathcal{B} (v) } z_\Phi[\pi_1,k];$$
\item \emph{For all spatial next}: $$\Psi=\forall_{\mathcal{B}} \bigcirc \Phi\rightarrow z_\Psi[v,k] = \bigwedge_{\pi \in \lambda^\mathcal{B} (v)} z_\Phi[\pi_1,k];$$
\item \emph{There exists spatial until}: $$\Psi=\exists_{\mathcal{B}} \Phi_1 \mathcal{U}_\kappa \Phi_2\rightarrow$$ $$z_\Psi[v,k]=\underset{\pi\in\lambda^{\mathcal{B}}(v), i\in(0,\kappa]}\bigvee(z_{\Phi_2}[\pi_i,k]\wedge  \underset{j\in[0,i)}\bigwedge z_{\Phi_1}[\pi_j,k]);$$
\item \emph{For all spatial until}: $$\Psi=\forall_{\mathcal{B}} \Phi_1 \mathcal{U}_\kappa \Phi_2\rightarrow$$ $$z_\Psi[v,k]=\underset{\pi\in\lambda^{\mathcal{B}}(v),i\in(0,\kappa]}\bigwedge(z_{\Phi_2}[\pi_i,k]\wedge  \underset{j\in[0,i)}\bigwedge z_{\Phi_1}[\pi_j,k]);$$
\item \emph{Temporal eventually}: $$\Psi=F_{[k_1\Delta t,k_2\Delta t)} \Phi\rightarrow z_\Psi[v,k] = \bigvee_{k^\prime=k_1,\cdots,k_2} z_\Phi[v,k^\prime];$$
\item \emph{Temporal always}: $$\Psi=G_{[k_1\Delta t,k_2\Delta t)} \Phi\rightarrow z_\Psi[v,k] = \bigwedge_{k^\prime=k_1,\cdots,k_2} z_\Phi[v,k^\prime];$$
\item \emph{Temporal until}: 
$$\Psi=\Phi_1 U_{[k_1\Delta t,k_2\Delta t)} \Phi_2\rightarrow$$ $$z_\Psi[v,k] = \underset{k^\prime=k_1,\cdots,k_2}\bigvee (z_{\Phi_2}[v,k^\prime]  \wedge  \underset{k^{\prime\prime}=k_1,\cdots,k'}\bigwedge z_{\Phi_1}[v,k^{\prime\prime}]).$$
\end{itemize}

Note that $z_\Psi[v,k] \in [0,1]$ is not required to be declared an integer since it is automatically enforced to take binary values. Finally, the problem of satisfying a general SpaTeL formula, $Q_0 \models \Phi$, reduces to the following constraint:
\begin{equation}
z_\Phi(v_\iota,0)=1,
\end{equation}
where $v_\iota$ is the root node of quad transition system.

\subsection{Robustness-Based Encoding}
In this section, we briefly explain how to incorporate SpaTeL robustness into the mixed-integer encoding. The method is much in spirit of the method in \cite{sadraddini2015robust}, where the authors characterize the changes in the satisfaction of the specification with respect to the changes in the predicates. For a predicate in the form of $\sigma=(\mu \sim c )$, it is straightforward to see from \eqref{eqn:quantitative_semantics} and \eqref{eqn:quantitative_semantics_2} that $\frac{\partial\rho(\Phi,Q_0)}{\partial c} \in \{0,1\}$ (non-decreasing) or $\frac{\partial\rho(\Phi,Q_0)}{\partial c} \in \{0,-1\}$ (non-increasing), depending on the operators preceding the predicate. Therefore, by increasing (decreasing) the value of $c$ for a non-increasing (non-decreasing) predicate, a constraint is \emph{tightened}. Therefore, we alter the values of $c$ in the predicates as follows:  
\begin{equation}
\left \{ 
\begin{array}{ll}
c \gets c+\varrho &~ ~\frac{\partial\rho(\Phi,Q_0)}{\partial c} \in \{0,-1\}, \\
c \gets c-\varrho &~ ~\frac{\partial\rho(\Phi,Q_0)}{\partial c} \in \{0,1\}.
\end{array}
\right.
\end{equation}
Next, we add the constraint $\varrho \ge 0$ to ensure satisfaction of $\Phi$. It is easy to show that the maximum $\varrho$ that renders $Q_0 \models \Phi$ is equal to $\rho(\Phi,Q_0)$.

\subsection{High Level Planning}  
\label{sec:LP}
In the previous sections, we formulated the dynamics and SpaTeL objectives as mixed-integer constraints. We formulate the discrete-time version of Problem 1 as the following MILP:
\begin{equation}
\begin{array}{ccl}
\label{eqn:MILP}
\underset{k=0,1,\cdots,K} {f[k]}  = & argmin &  -\alpha~ \varrho+ \mathcal{J}_f(\mathcal{N}[K]) \\&&  + \sum_0^K \mathcal{J}_r(\mathcal{N}[k],f[k]),
\vspace{5pt}
\\
& s.t. & \mathcal{N}[k+1]=\mathcal{F}(\mathcal{N}[k], f[k]), \\
& & z_\Phi(v_\iota,0)=1, \\&&
\varrho \ge 0,

\end{array}
\end{equation}
where $\mathcal{J}_f$ and $\mathcal{J}_r$ are the discrete time versions of the endpoint and running cost, respectively. Note that we assume the costs are linear functions. In this paper, we are particularly interested in the following cost:
\begin{equation}
J_r(\mathcal{N}[k],f[k])= \sum f[k],
\label{equ:movements}
\end{equation}
which corresponds to the total number of robot displacements (energy). Note that all the values of $f$ are non-negative. 

In case the MILP above is infeasible, no control strategy is able to satisfy the SpaTeL formula. In this case, we relax the last constraint $\varrho \ge 0$, and choose a very large value for $\alpha$ (or remove the other costs). Therefore, the resulting solution solely maximizes the SpaTeL robustness, which is a negative value. In other words, the SpaTeL violation is minimized.

%

\subsection{Low Level Control Policy}
\label{sec:control}
The decision variables $f_{[i,j]}^{[i',j']}[k]$ are obtained from the solution to \eqref{eqn:MILP}. The only remaining problem is to choose the set of individual robots that must be moved from cell $[i,j]$ to adjacent cells. For this purpose, we choose $f_{[i,j]}^{[i',j']}[k]$ number of robots that are closest to the edge between $[i,j]$ and $[i',j']$ and move them with a constant velocity on a straight line. In other words, if $\mathcal{R}_{[i,j]}[k]$ is the the set of robot indices that are located inside cell $[i,j]$ at time step $k$ and $\mathcal{R}_{[i,j]}^{[i',j']}[k]$ is the set of robot indices that are supposed to be moved from $[i,j]$ to $[i',j']\in\Omega([i,j])$ at time step $k$, the control law would be:
\begin{equation}
\begin{array}{r}
u_r(t)=\frac{a}{2^D\Delta t}.\left(
\begin{array}{c}
j'-j \\ i-i'
\end{array}
\right),~~~~~
r\in\mathcal{R}_{[i,j]}^{[i',j']}[k],\\ k\Delta t \leq t < (k+1)\Delta t.
\end{array}
\end{equation}
Algorithm \ref{alg:control} presents the procedure to determine $\mathcal{R}_{[i,j]}^{[i',j']}[k]$.

\begin{algorithm}
 \KwData{$[i,j]$ (cell index), $\mathcal{R}_{[i,j]}[k]$ (robots inside that cell), $f[k]$ (flow variables)}
 \KwResult{$\mathcal{R}_{[i,j]}^{[i',j']}[k]$ ( set of robots that are moved from $[i,j]$ to a neighbor)}
 \While{$\sum_{[i',j']\in\Omega([i,j])} f_{[i,j]}^{[i',j']}[k]> 0$}{
  find robot $r\in\mathcal{R}_{[i,j]}[k]$ that has the minimum distance to the edges of $[i,j]$\;
  add $r$ to $\mathcal{R}_{[i,j]}^{[i',j']}[k]$ where $[i',j']$ is the neighbor of $[i,j]$ that $r$ was closest to\;
  remove $r$ from $\mathcal{R}_{[i,j]}[k]$\;
  $f_{[i,j]}^{[i',j']}[k]\leftarrow f_{[i,j]}^{[i',j']}[k]-1$\;
 }
 \caption{How to assign robots to move from a cell to its neighbors}
 \label{alg:control}
\end{algorithm}

\begin{remark}
As mentioned earlier, we do not consider physical sizes for robots in this paper. In practice, careful strategies are required for collision avoidance among the robots. This issue will be further investigated in future work. As a preliminary solution, we propose the following approach. Let $n^{cap}$ be the maximum number of robots that can populate one cell without physically overlapping one another. We basically add the specification that for all cells the number of robots should not exceed $n^{cap}$, which guarantees there is always enough empty space for swarm movements. Localized policies such as the methods in \cite{van2011reciprocal} can be used for guaranteeing collision avoidance. 
\end{remark}

\begin{figure*}[t]
\begin{center}
\begin{tabular}{cccccc}
\includegraphics[width=0.142\textwidth]{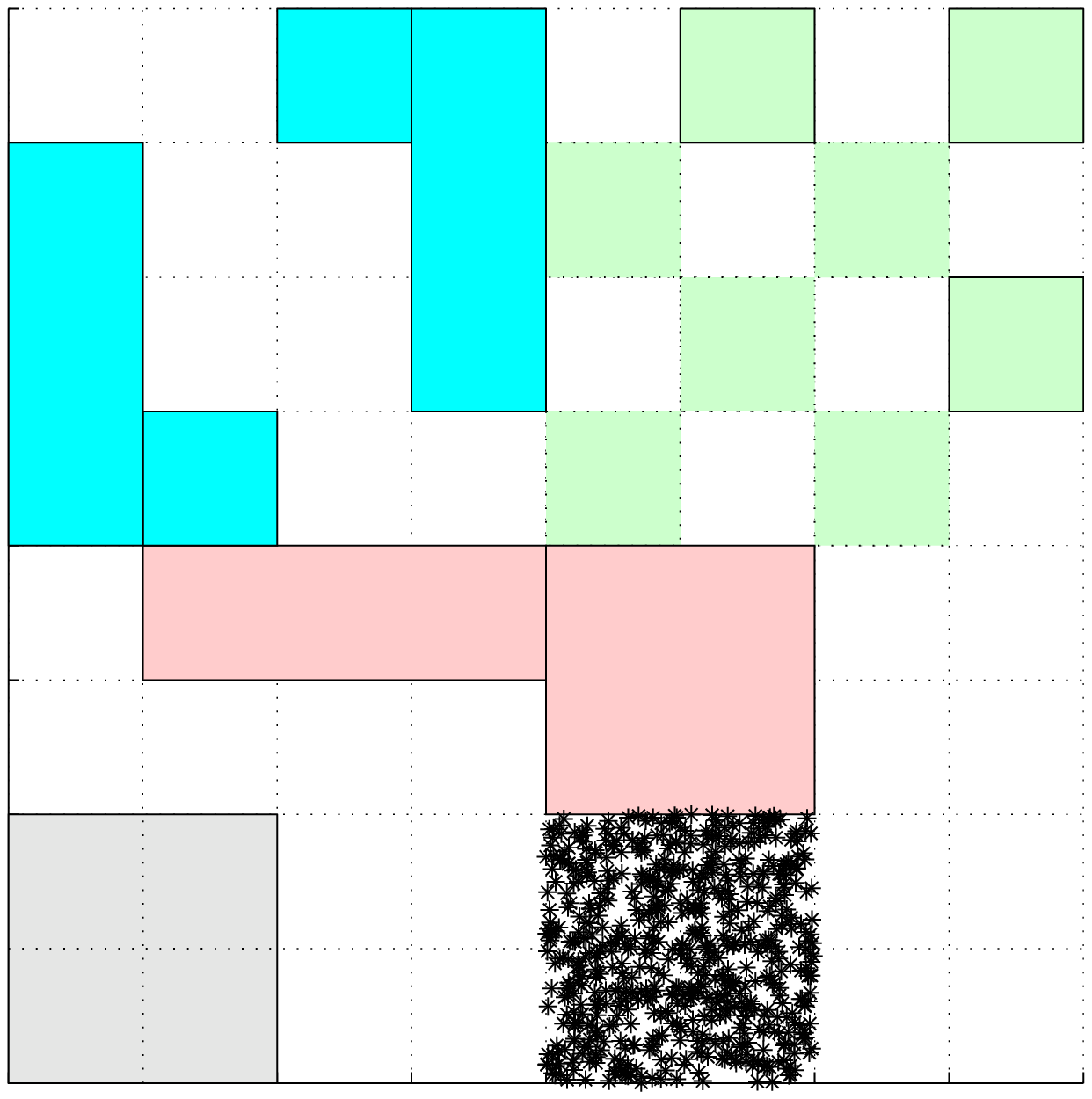} &
\includegraphics[width=0.142\textwidth]{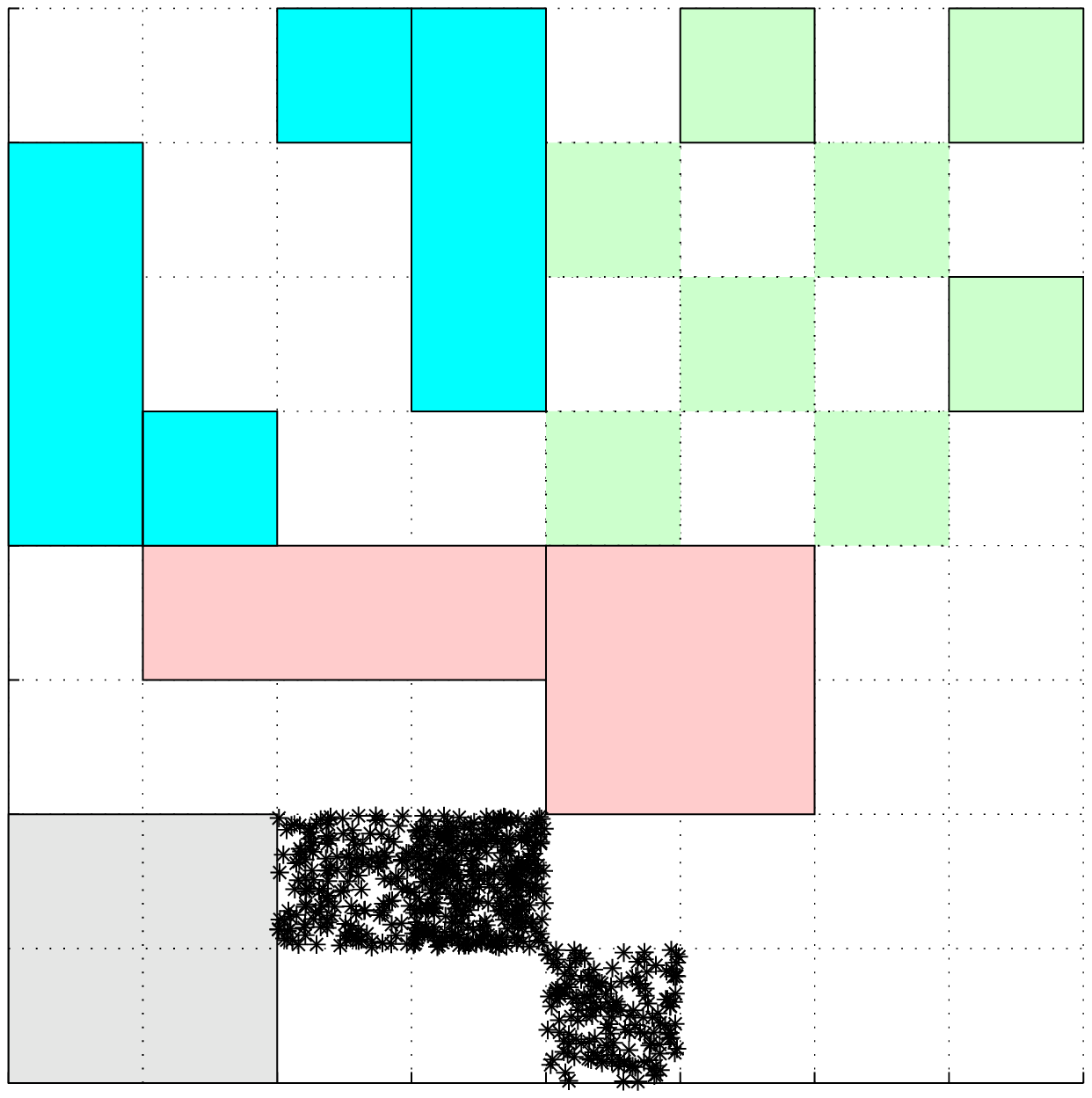} &  
\includegraphics[width=0.142\textwidth]{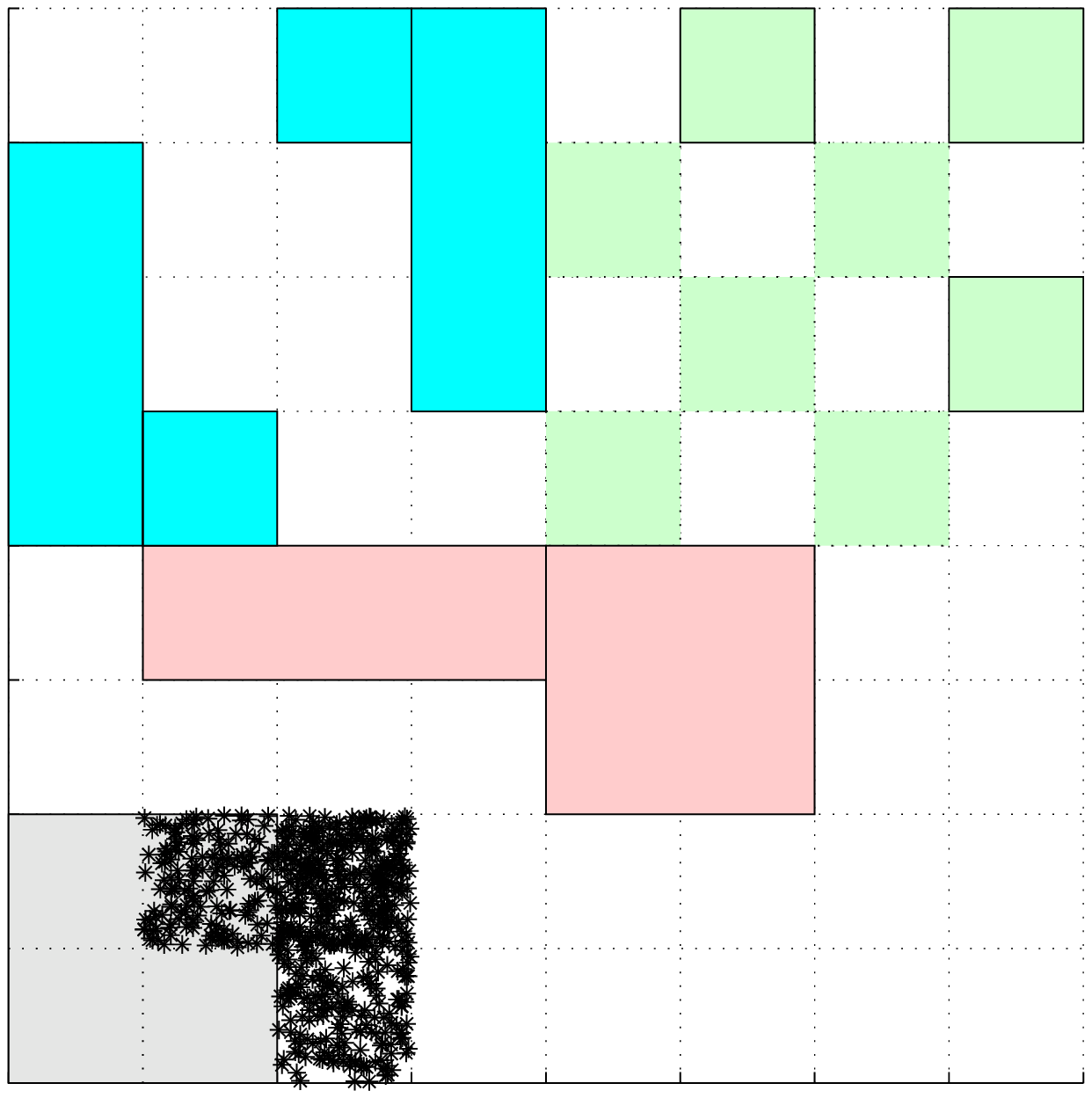} &  
\includegraphics[width=0.142\textwidth]{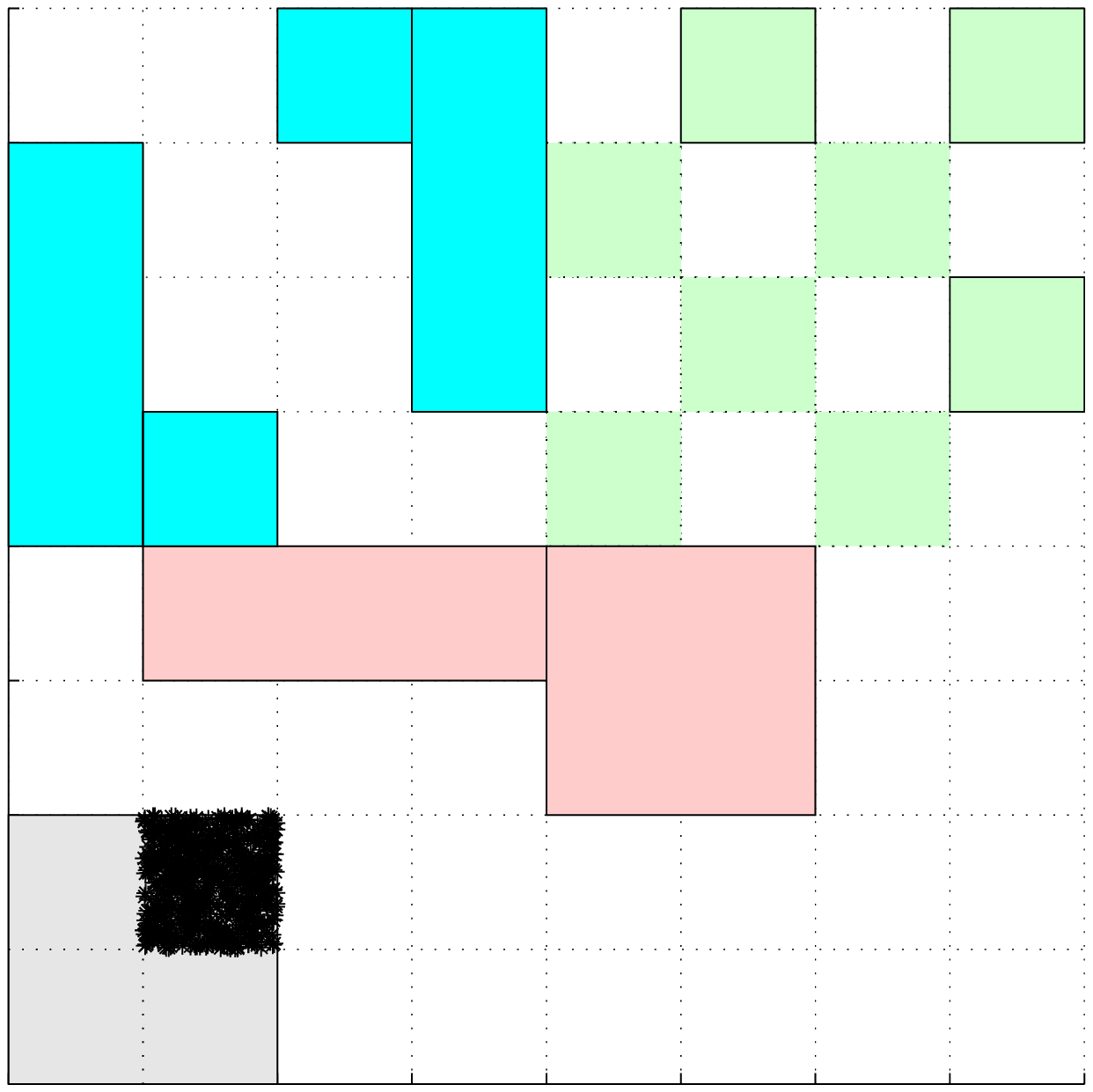} &
\includegraphics[width=0.142\textwidth]{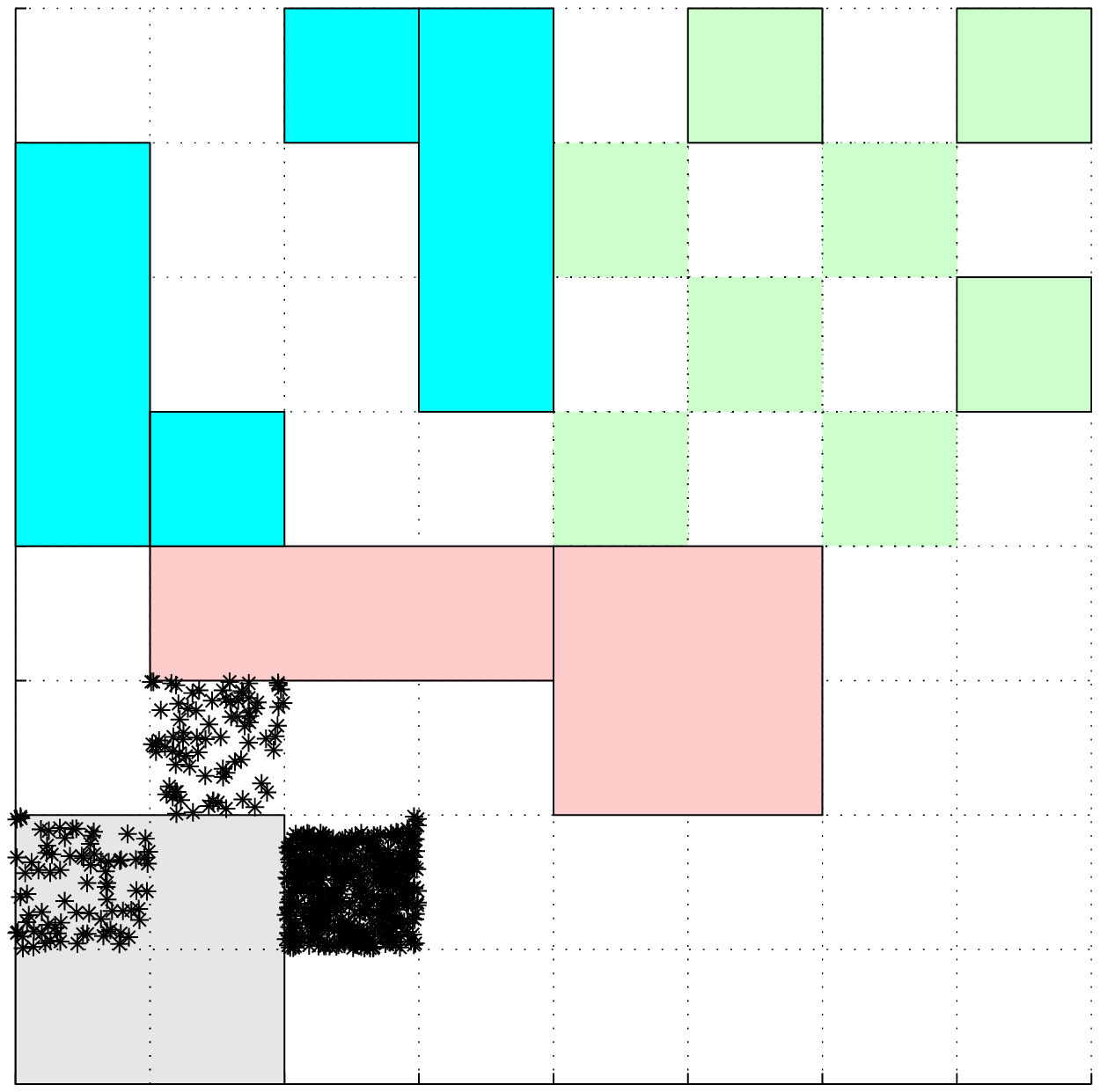} &
\includegraphics[width=0.142\textwidth]{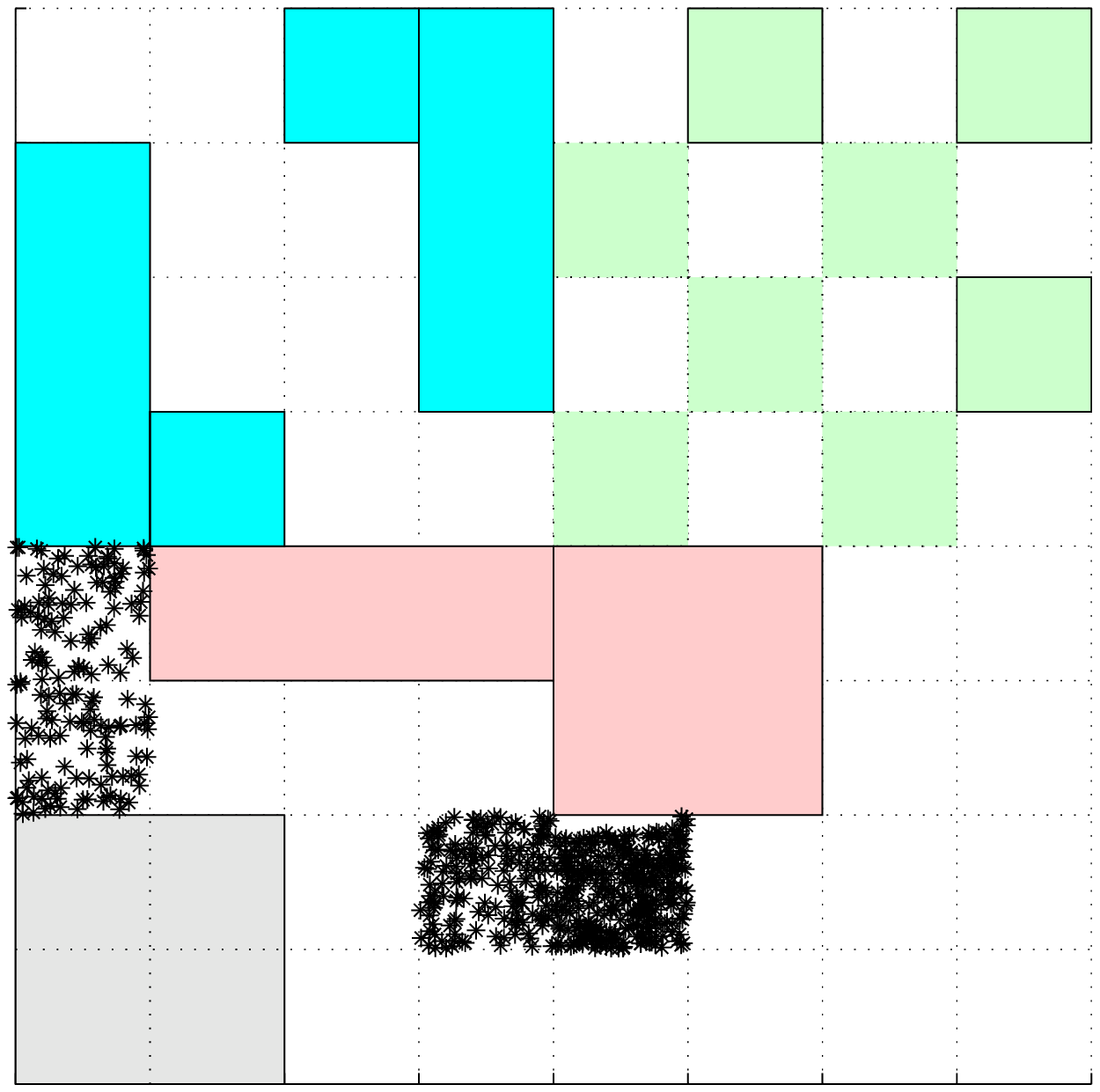}
\\ 
a) $t=0$ & b) $t=3$ & c) $t=7$ & d) $t=11$ & e) $t=16$ & f) $t=18$ \\
\includegraphics[width=0.142\textwidth]{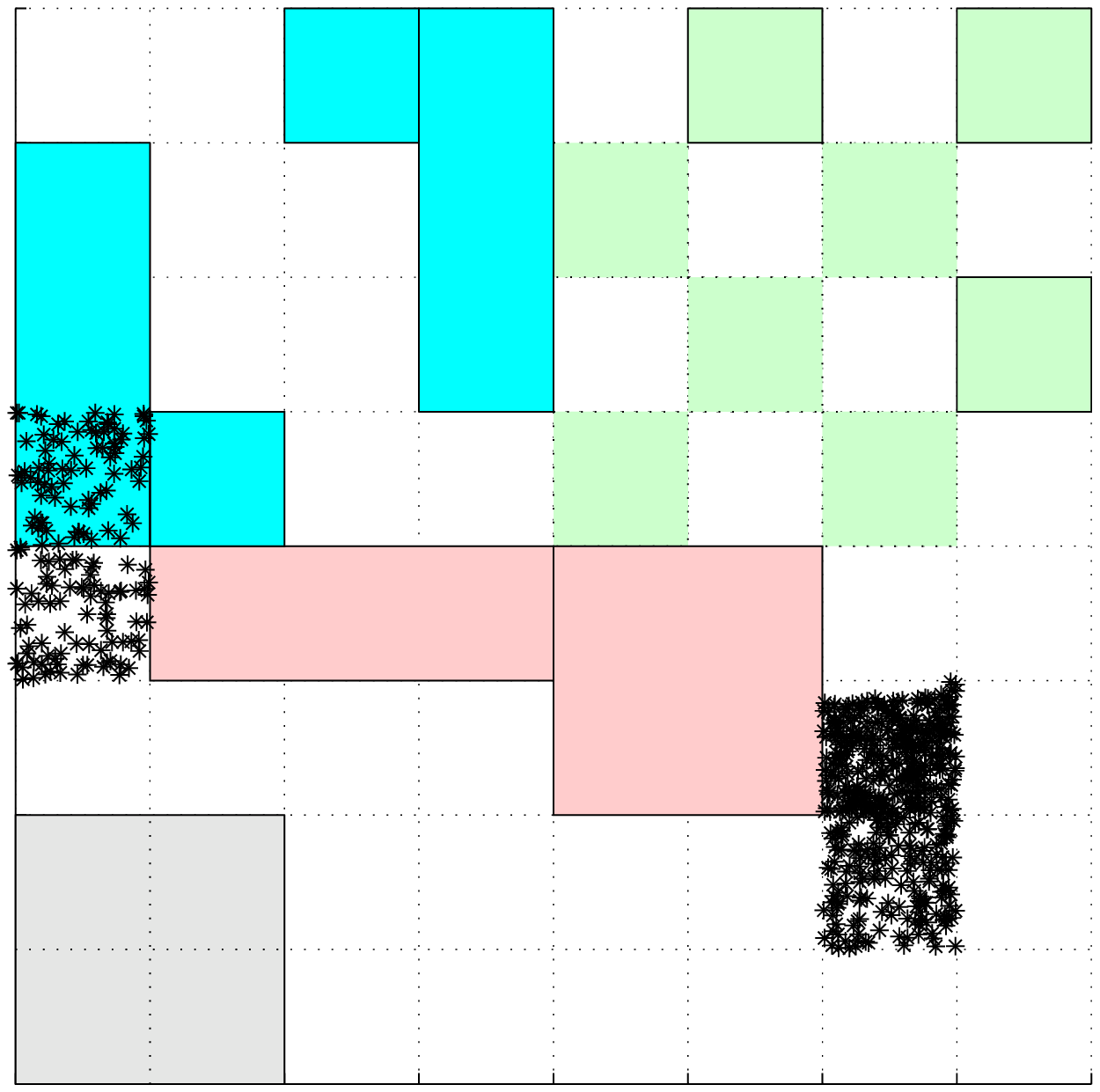} &
\includegraphics[width=0.142\textwidth]{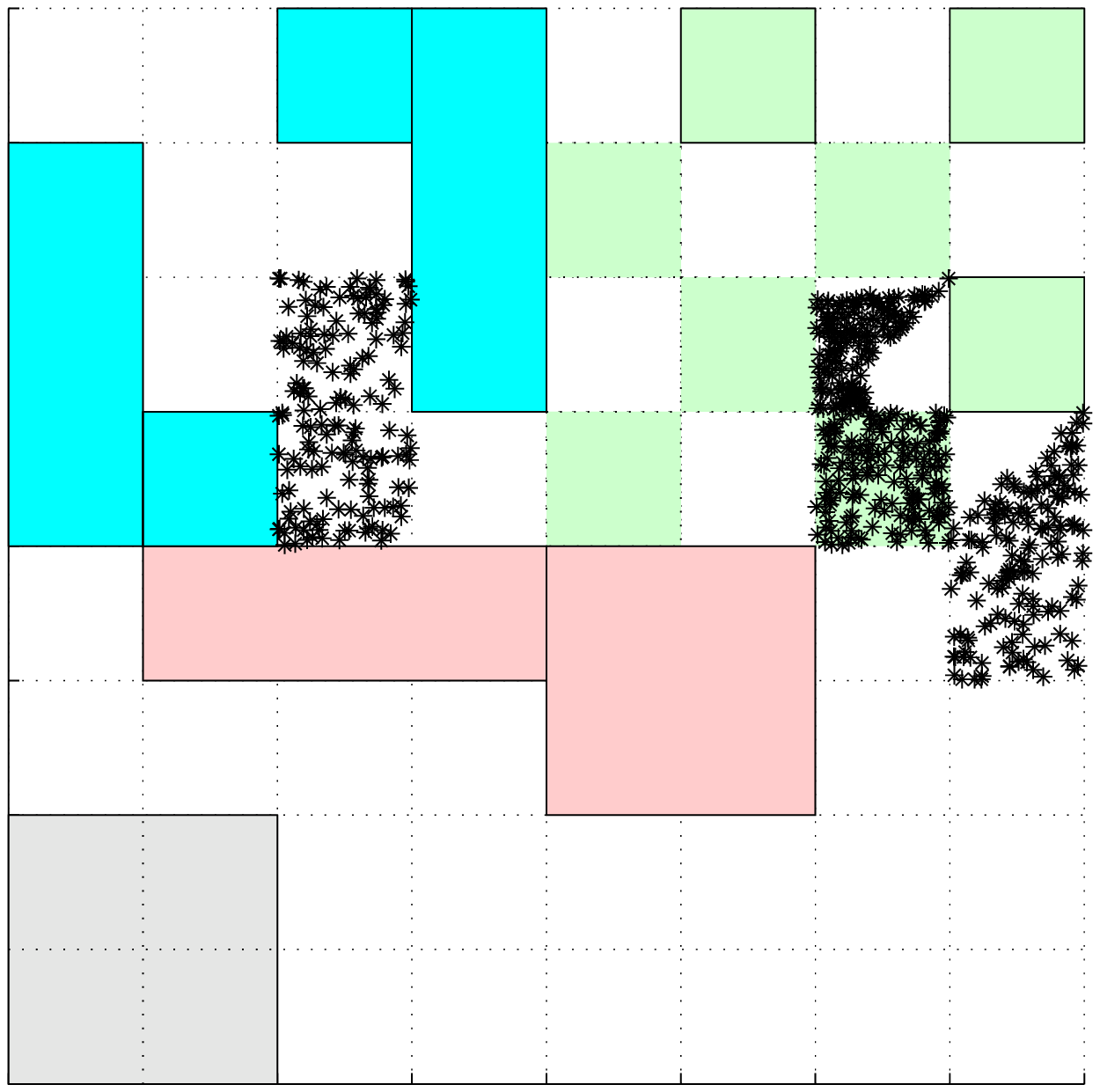} &
\includegraphics[width=0.142\textwidth]{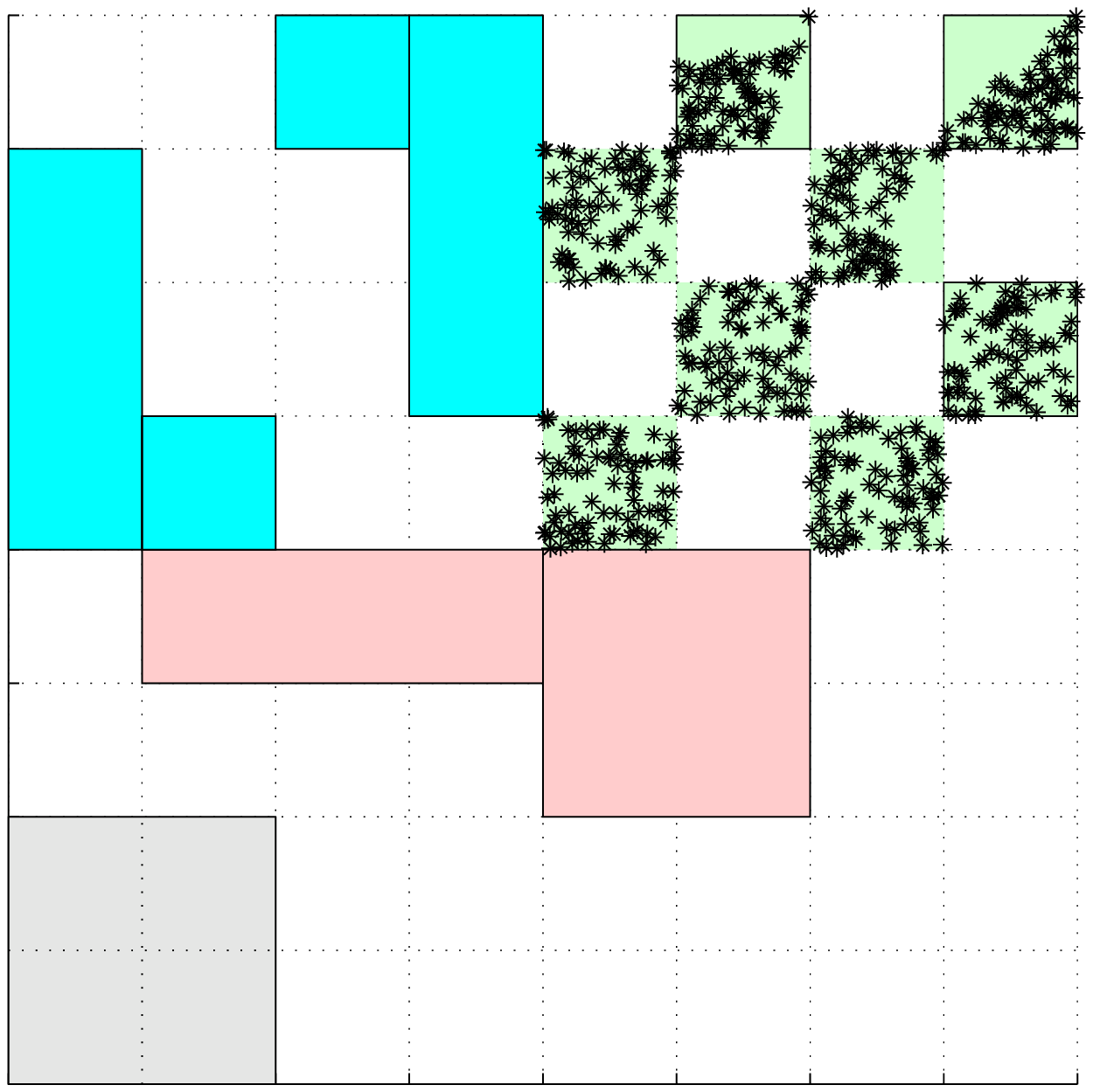} &  
\includegraphics[width=0.142\textwidth]{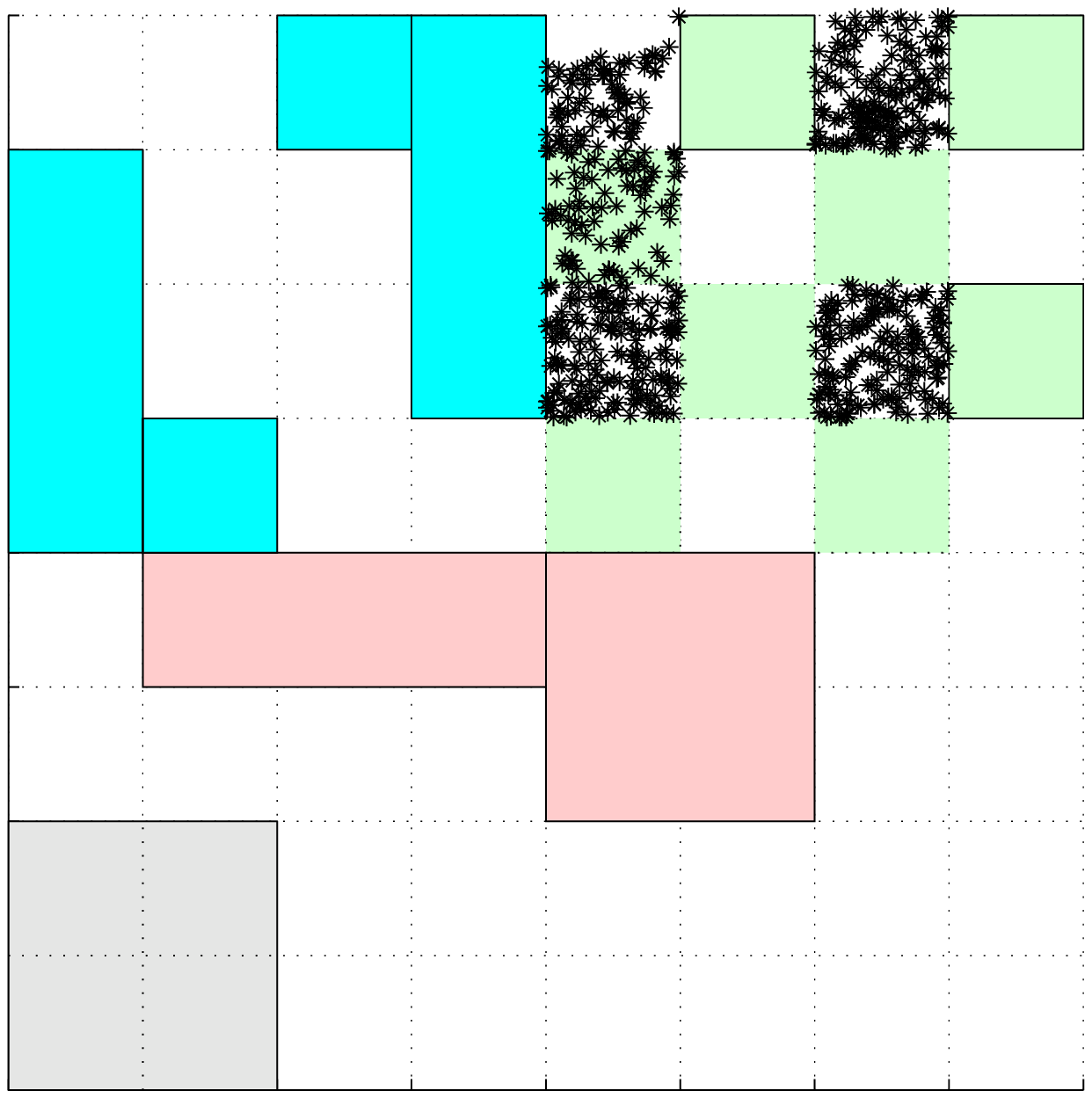} &
\includegraphics[width=0.142\textwidth]{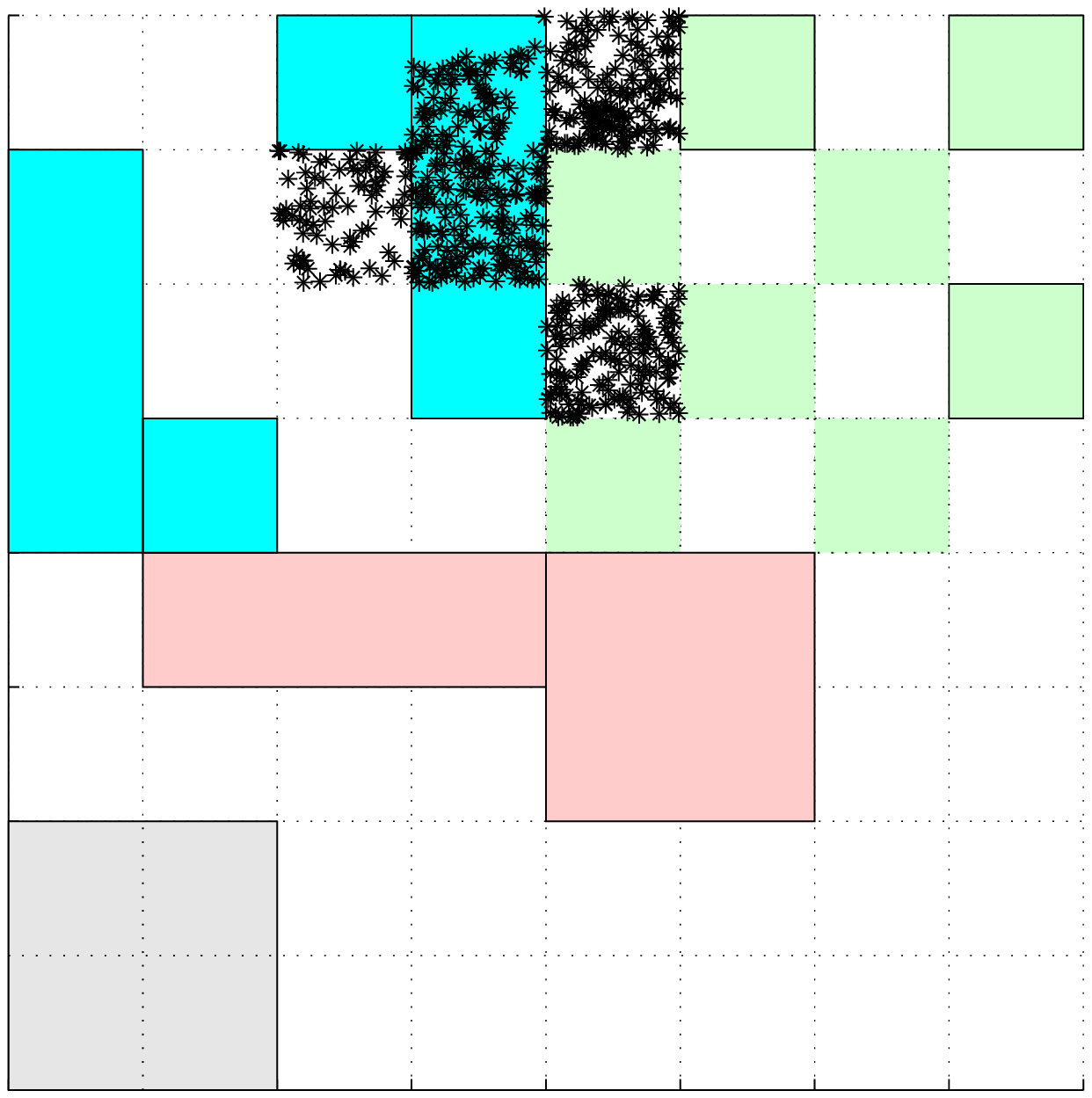} &
\includegraphics[width=0.142\textwidth]{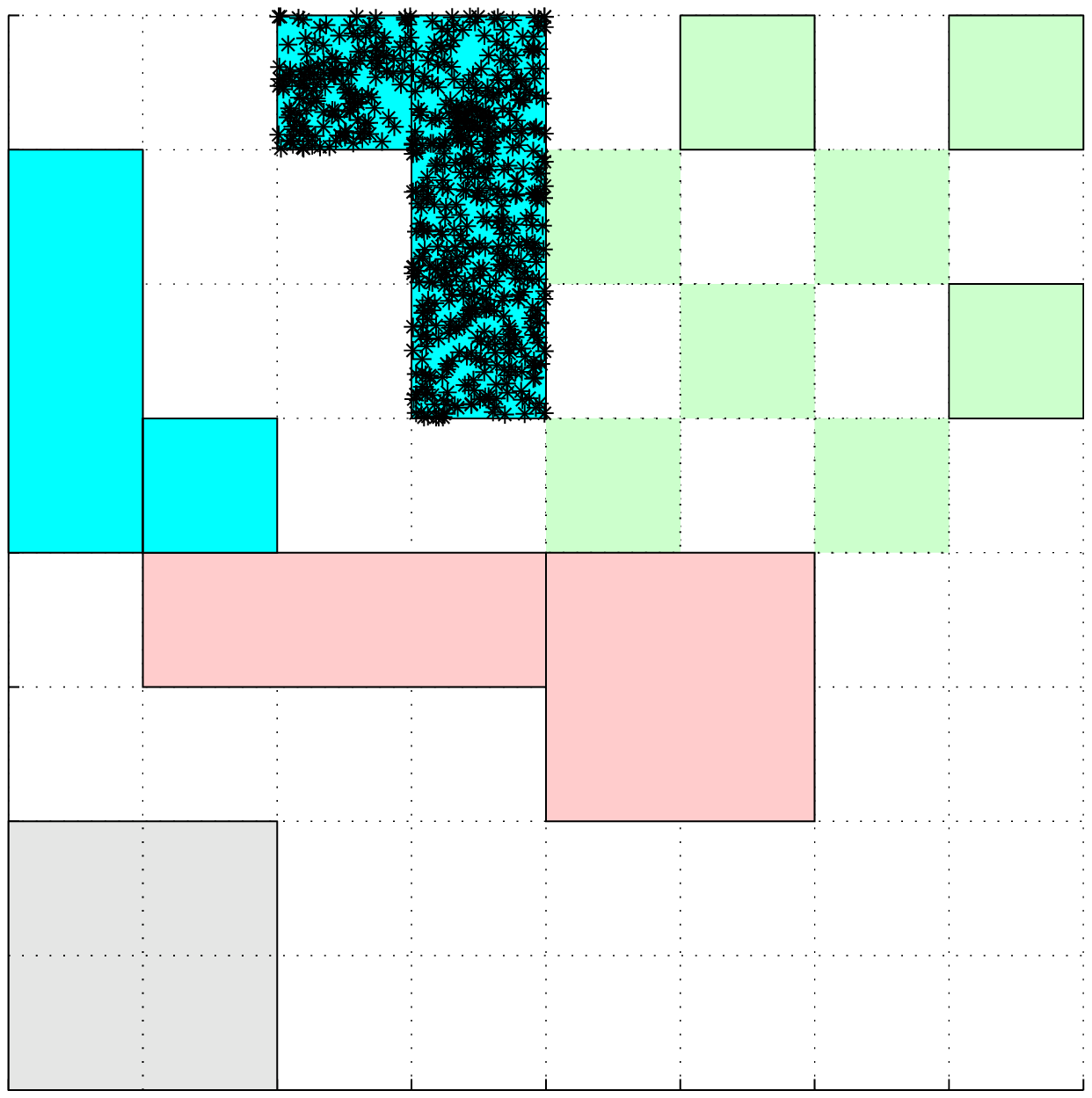}  
\\ 
g) $t=21$ & h) $t=24$ & i) $t=27$ & j) $t=30$ & k) $t=32$ & l) $t=40$
\end{tabular}
\caption{Case study: Snapshots of the optimal swarm movement satisfying SpaTel formula \eqref{eqn:case_spatel} starting from the initial condition shown in figure a). First the robots are gathered in one cell in the grey region to satisfy $\varphi_3$, then robots move toward forming the checkerboard pattern, satisfying $\varphi_2$. Finally robots move to the populate the upper L-shaped pattern, satisfying   $\varphi_5$.}
\label{fig:case1}
\end{center}
\end{figure*} 
\begin{figure*}[t]
\begin{center}
\begin{tabular}{cccccc}
\includegraphics[width=0.142\textwidth]{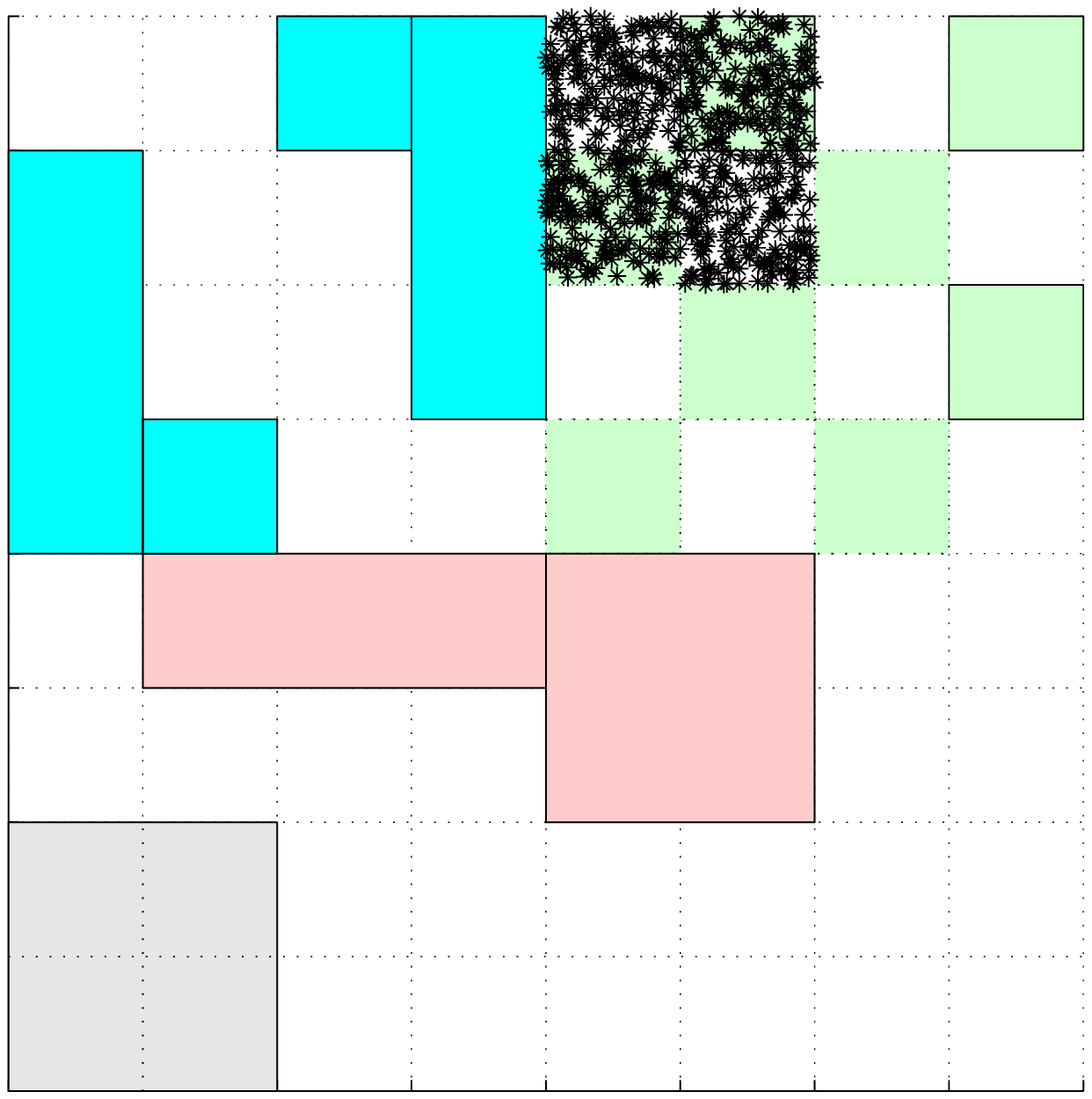} &
\includegraphics[width=0.142\textwidth]{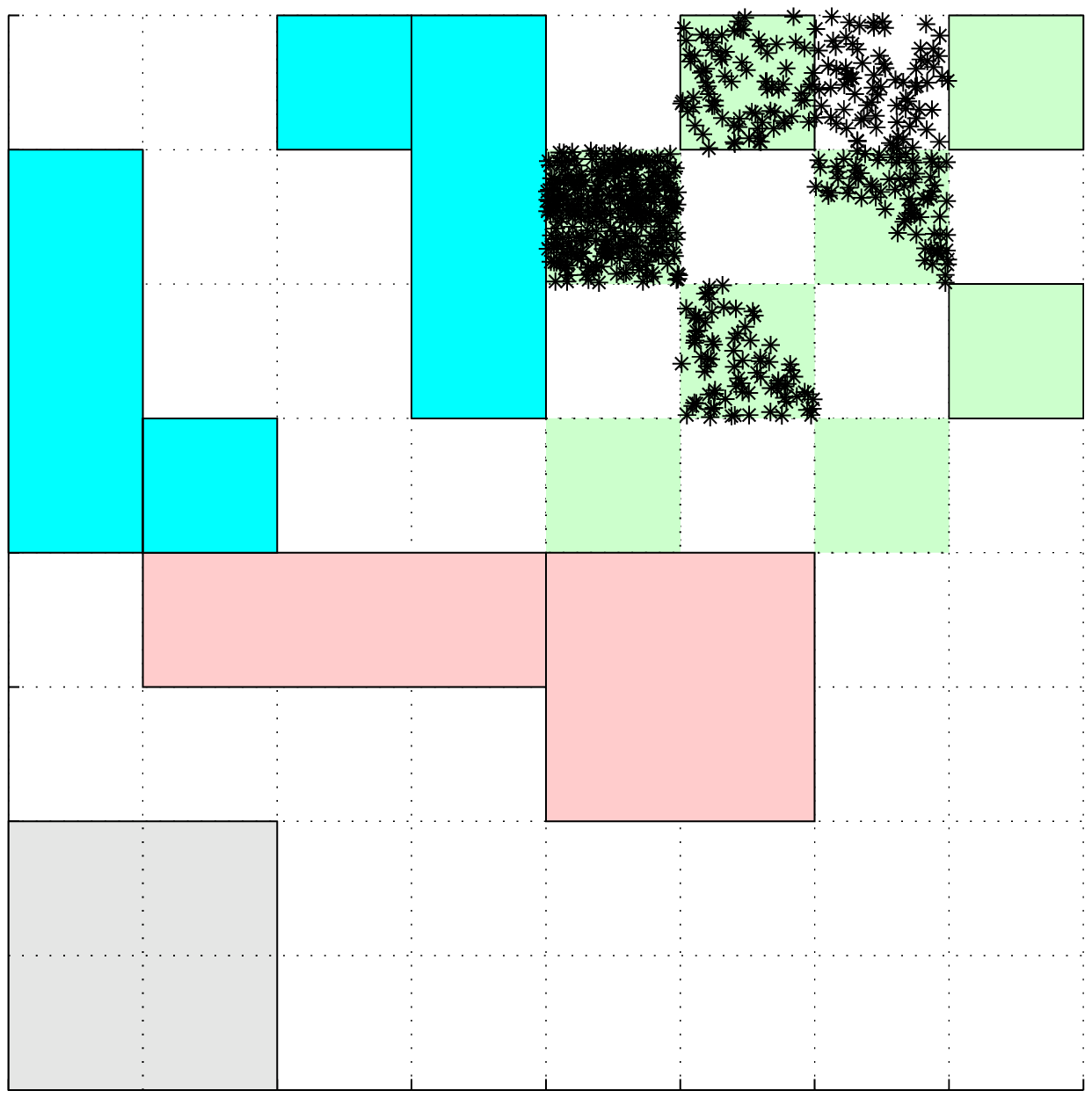} &  
\includegraphics[width=0.142\textwidth]{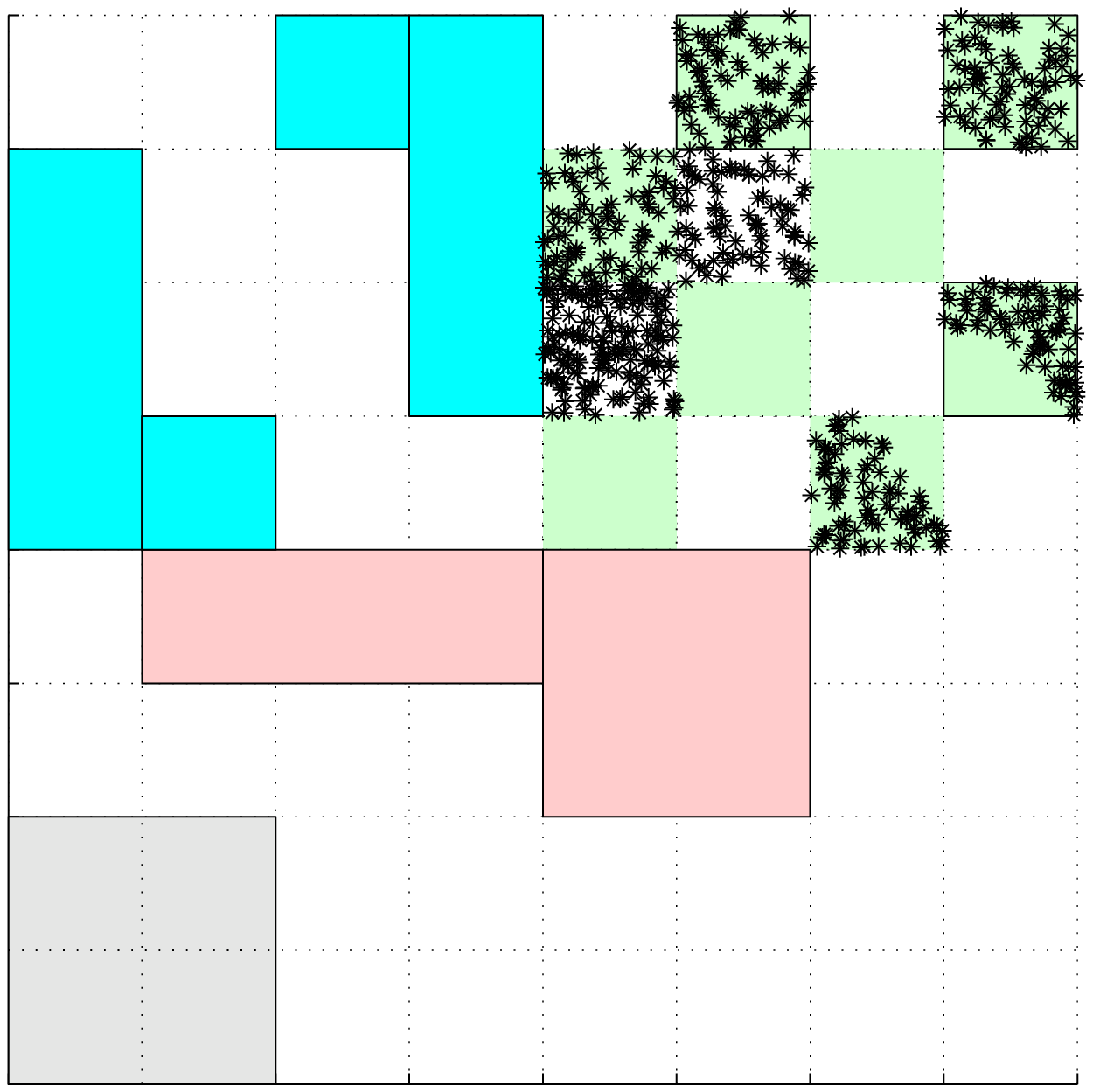} &
\includegraphics[width=0.142\textwidth]{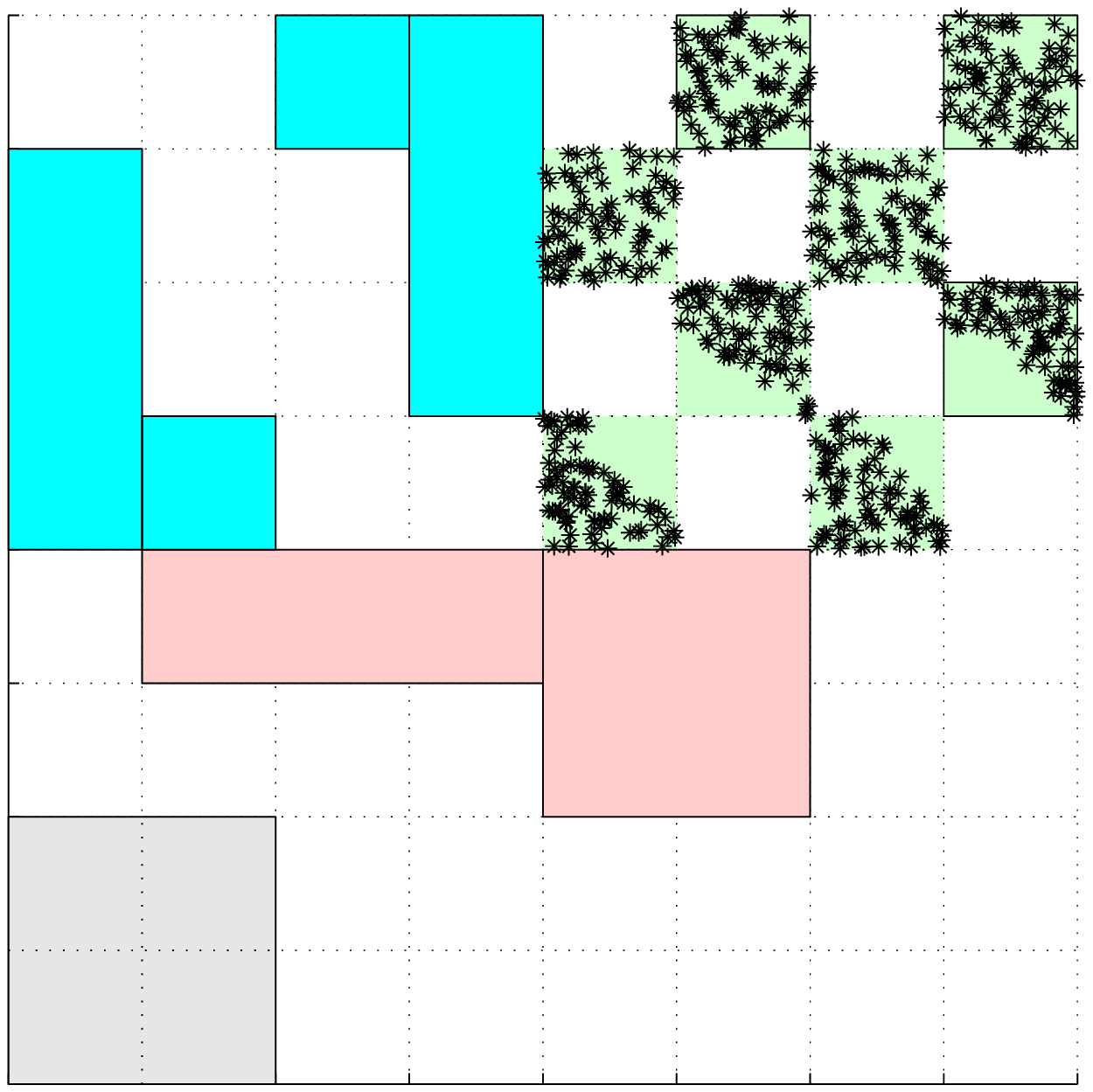} &
\includegraphics[width=0.142\textwidth]{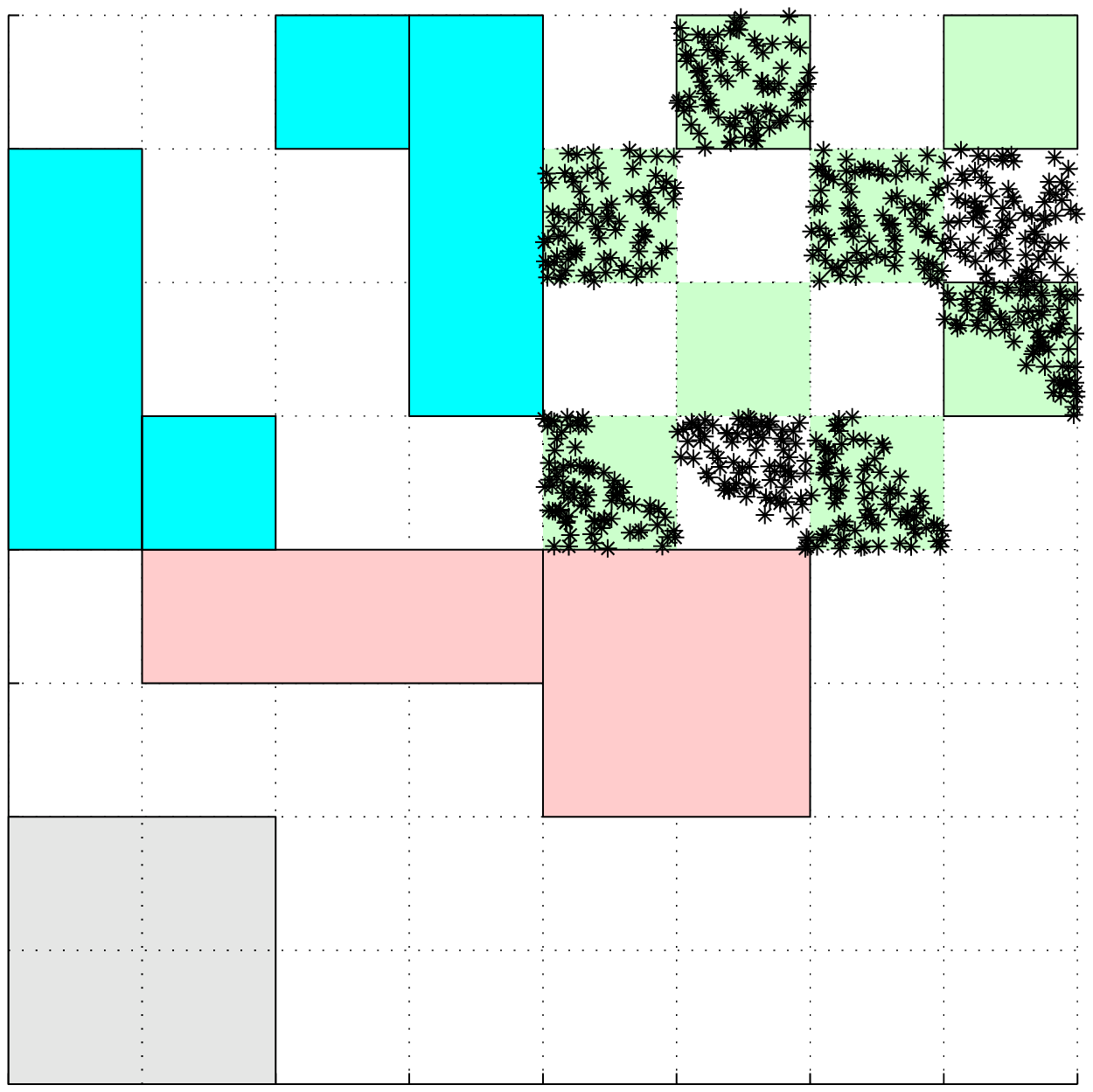} &
\includegraphics[width=0.142\textwidth]{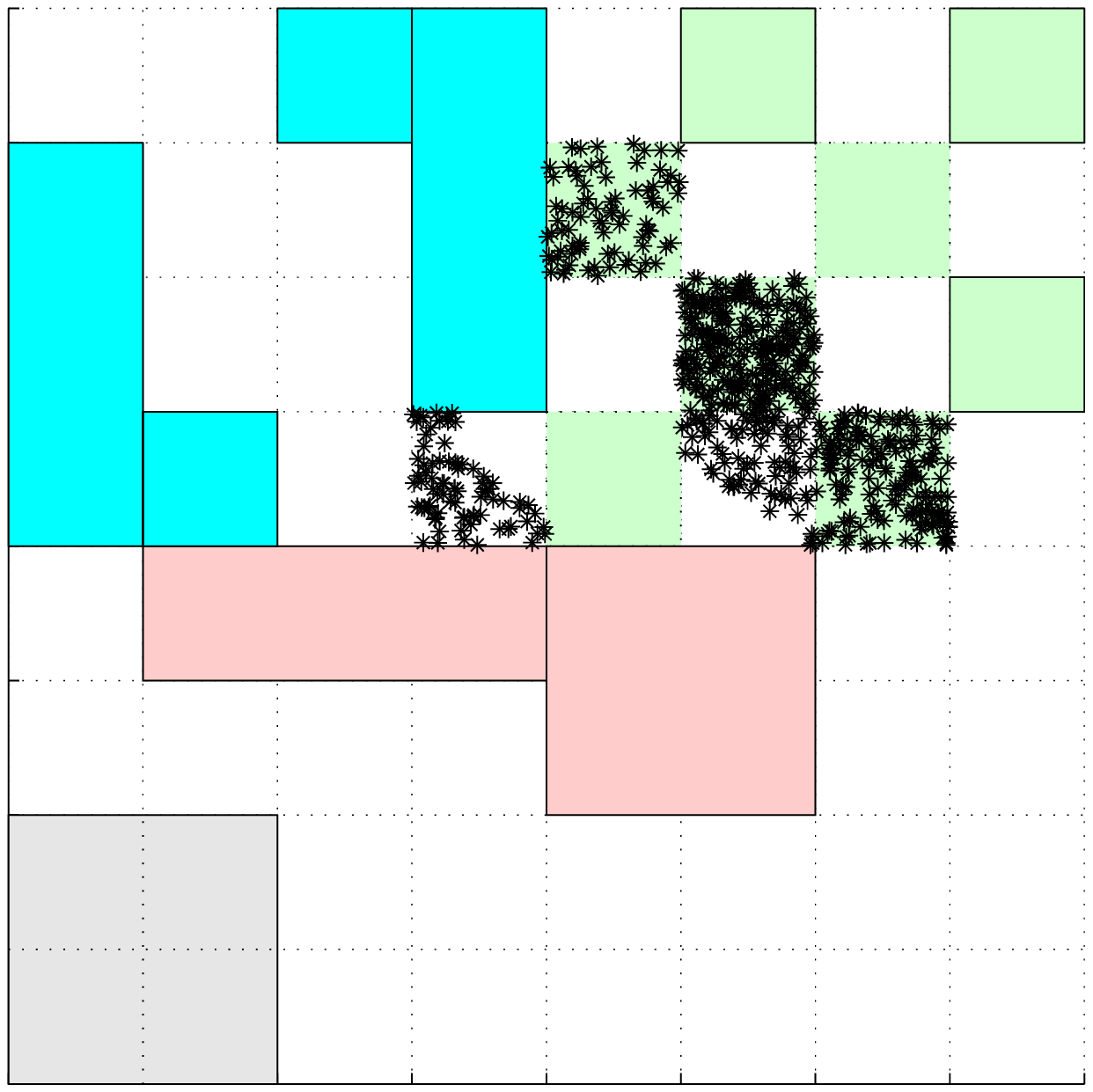}  
\\ 
a) $t=0$ & b) $t=1$ & c) $t=3$ & d) $t=6$ & e) $t=10$ & f) $t=13$ \\
\includegraphics[width=0.142\textwidth]{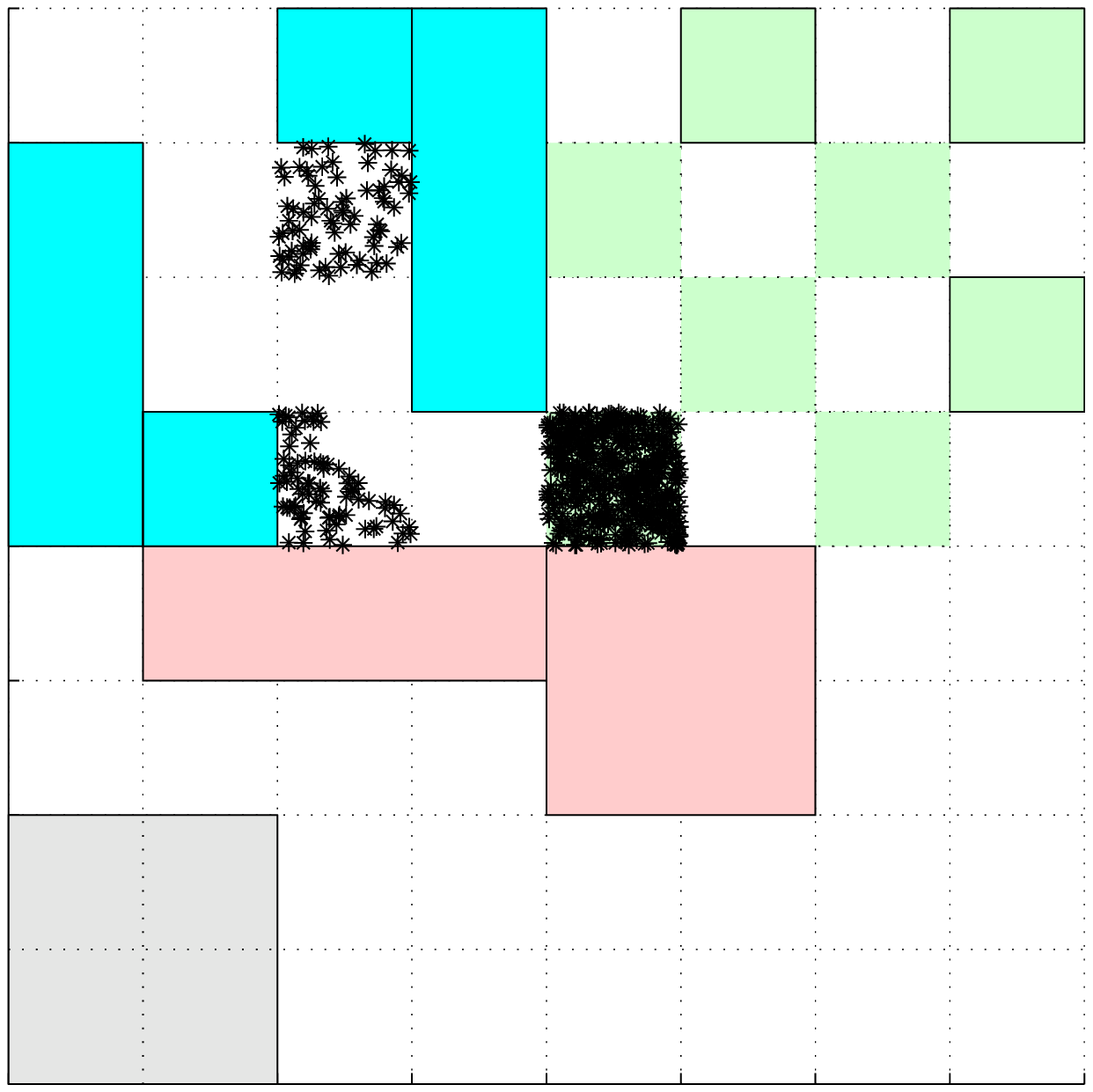} &
\includegraphics[width=0.142\textwidth]{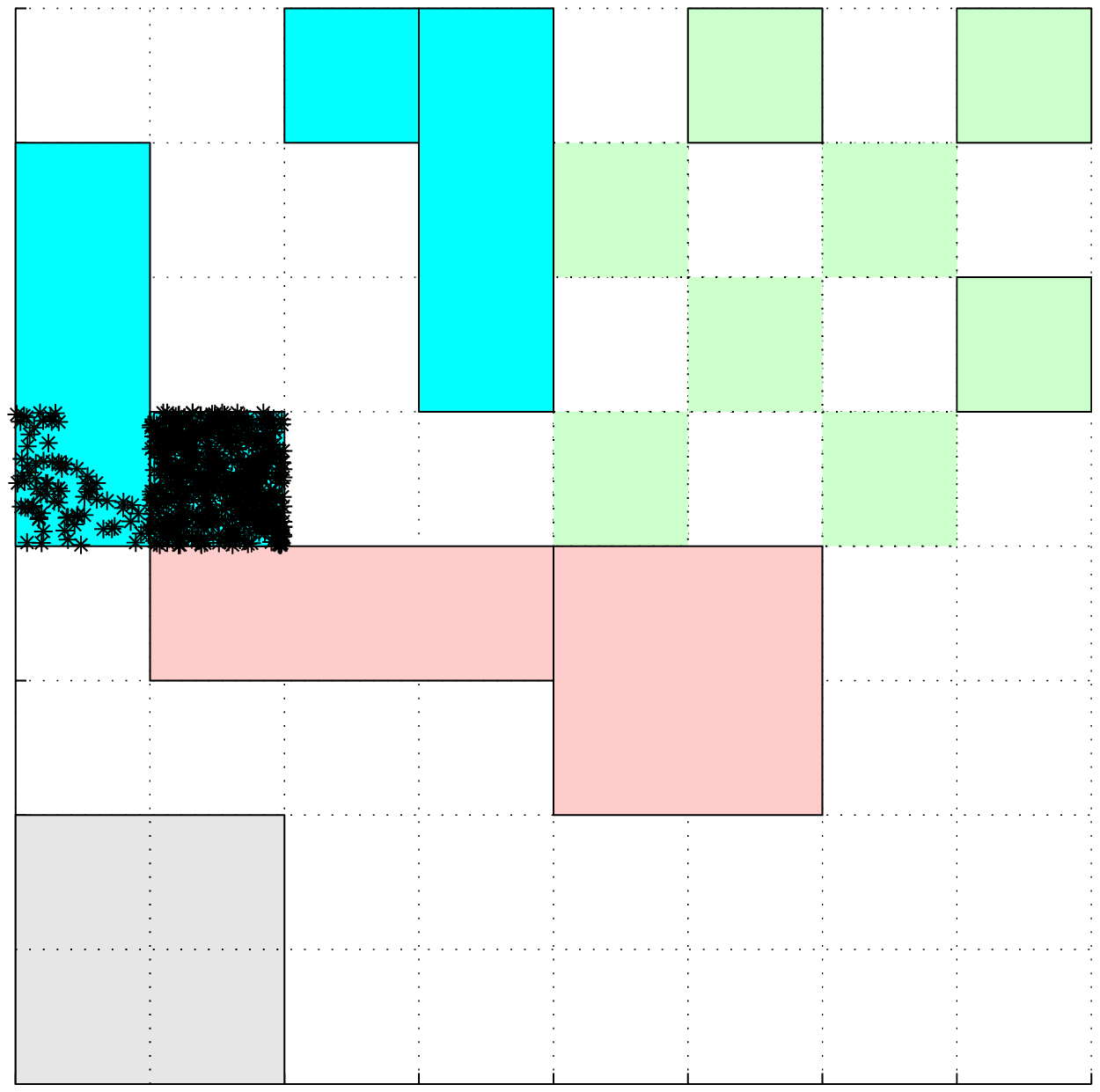} &  
\includegraphics[width=0.142\textwidth]{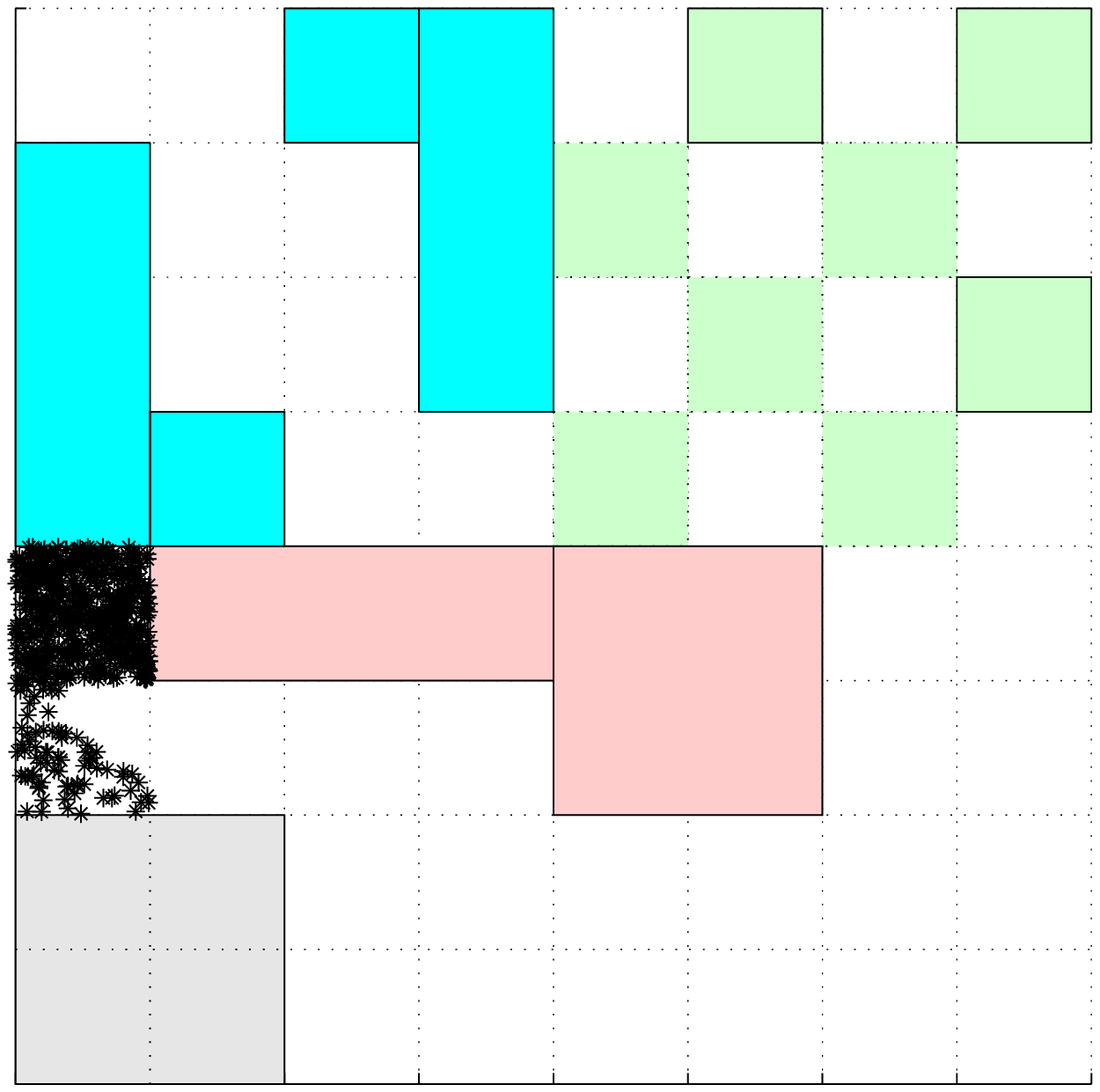} &
\includegraphics[width=0.142\textwidth]{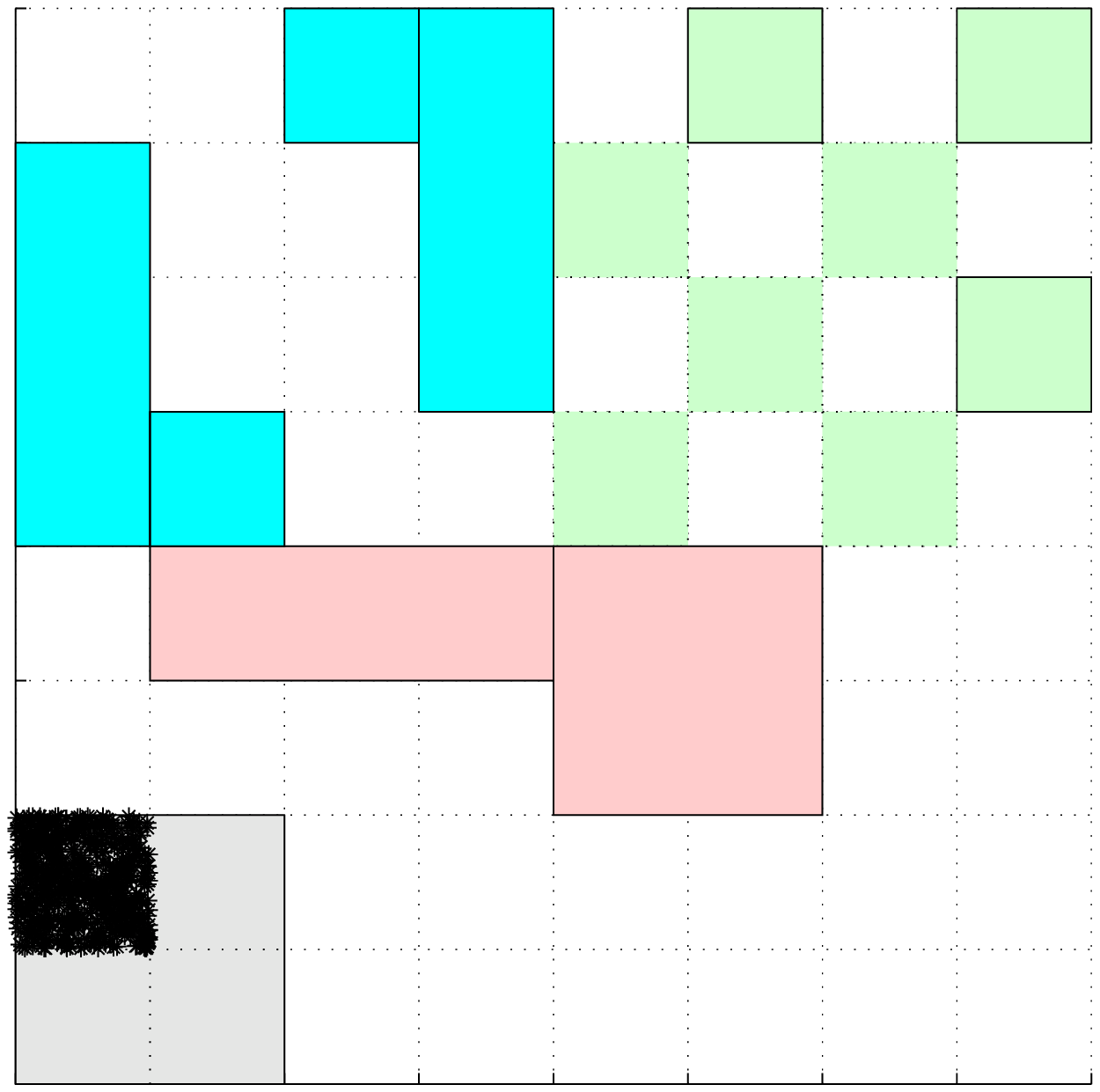} &
\includegraphics[width=0.142\textwidth]{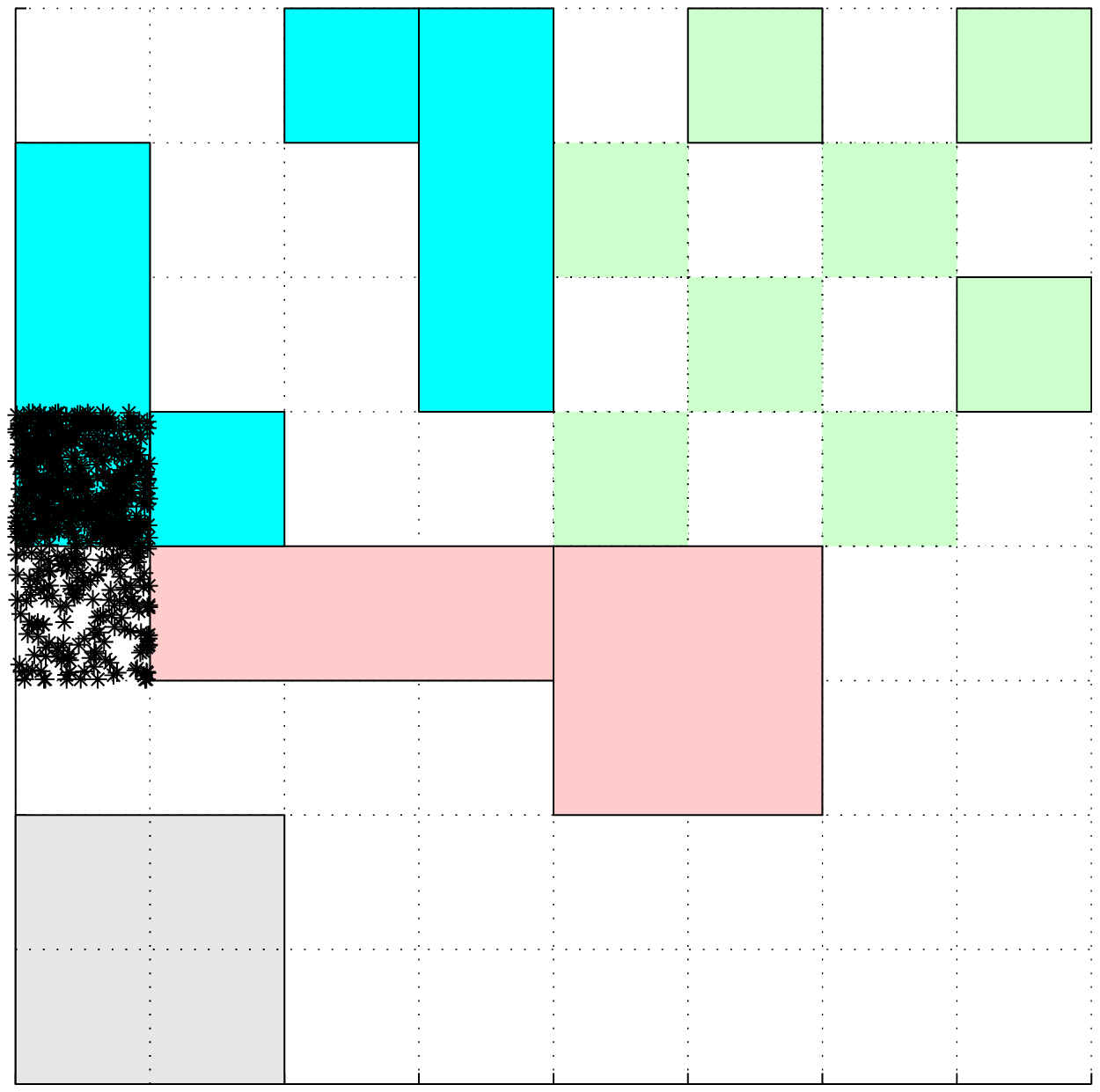} &
\includegraphics[width=0.142\textwidth]{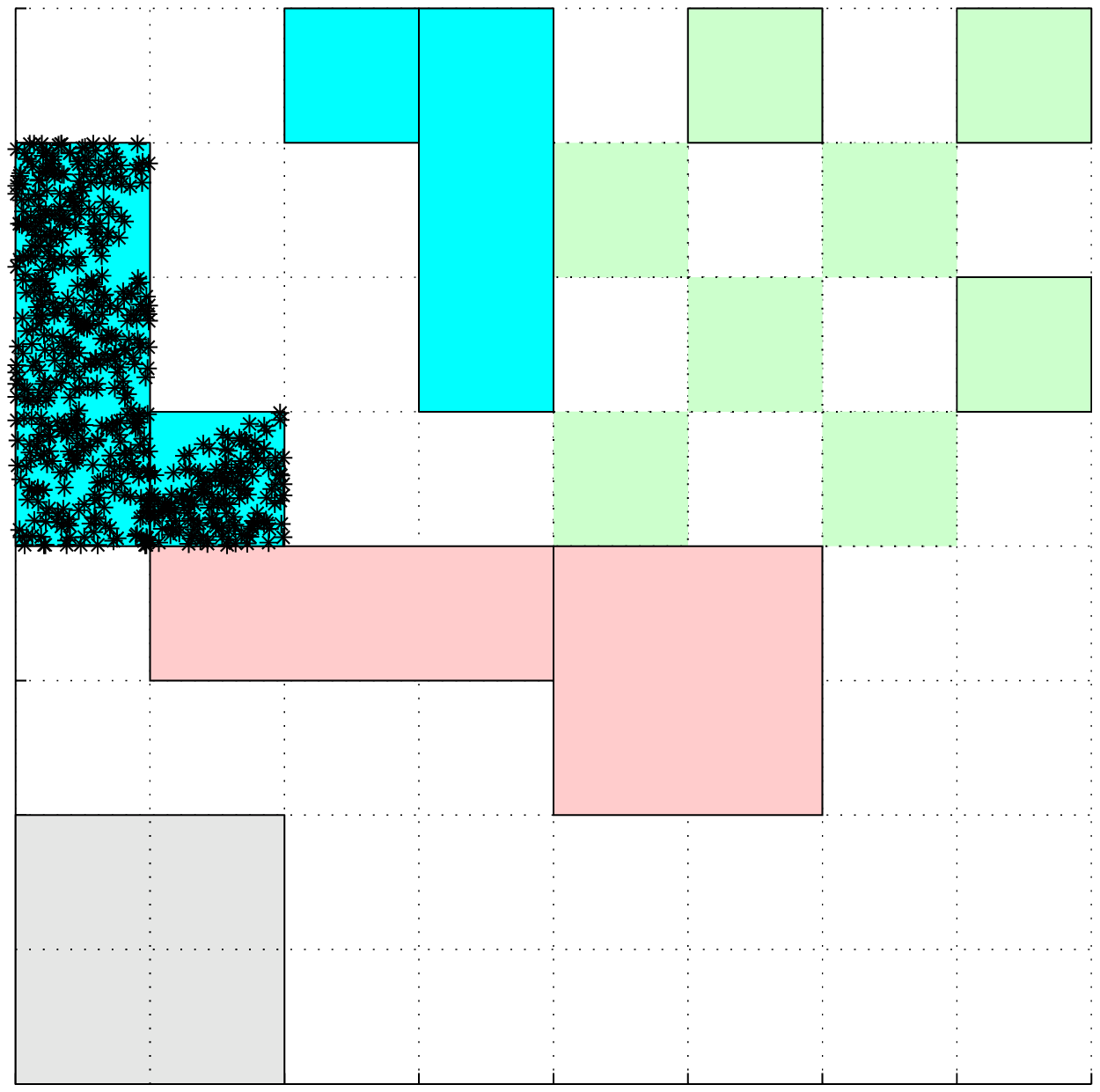}  
\\ 
g) $t=15$ & h) $t=18$ & i) $t=20$ & j) $t=22$ & k) $t=31$ & l) $t=40$ 
\end{tabular}
\caption{Case study: Snapshots of the optimal swarm movement satisfying SpaTel formula \eqref{eqn:case_spatel} starting from the initial condition shown in figure a). First the robots are forming the checkerboard pattern $\varphi_2$, then they gather in one grey cell to satisfy $\varphi_3$, and finally robots move to the populate the lower L-shaped pattern, satisfying $\varphi_6$.}
\label{fig:case2}
\end{center}
\end{figure*} 

\subsection{Complexity}
\label{sec:complexity}
The worst case complexity of the framework outlined in previous sections depends on the complexity of the MILP formulation in Sec. \ref{sec:mixed}. The complexity of a MILP problem grows exponentially, in the worst case, with respect to the number of integer variables and polynomially with respect to the continuous variables. It is worth to note that for large swarms, the flow variables can be approximated as continuous numbers and be rounded off after obtaining the MILP solution. This approximation significantly reduces the computational complexity without significantly altering the optimal solution and makes the complexity independent from the total number of robots. The number of integer variables in (\ref{eqn:MILP}) is $KP\left|\mathcal{V}_f\right|$, where $K$ is the total number of time steps, $P$ is the number of linear predicates in $\Phi$, and $|\mathcal{V}_f|$ is the number of cells in the gridded workspace. This suggests that the framework might not be scalable for grids with high resolutions or extremely complicated and long spatio-temporal requirements. However, as illustrated in the examples in Sec. \ref{sec:case}, quite complicated patterns are achievable in practice with relatively low computation time.

\section{Case Study}
\label{sec:case} 
$640$ robots in a workspace partitioned into a $8\times 8$ grid. The SpaTeL formula of \eqref{eqn:case_spatel} corresponding to the specification of Fig. \ref{fig:intro} is the target, where we set $\gamma_1=80,\gamma_2=640,\gamma_{3-6}=160$. The cost function that is minimized is the total number of robot displacements given by \eqref{equ:movements}. We demonstrate the results for two different initial conditions. A movie illustrating both cases is available on \url{https://youtu.be/x-uI8N9iN3I}.
\subsection{Case 1}
We set the initial configuration of robots to be in the uniformly distributed in the $SW$ quadrant of the $SE$ quadrant (see Fig. \ref{fig:case1} a). We formulate \eqref{eqn:MILP} as a MILP, which we solve using Gurobi \footnote{www.gurobi.com}. The MILP is solved in 54 seconds on a 3GHz Dual core Macbook Pro. Next, we move the robots according to the plan obtained from the solution of the MILP. Fig. \ref{fig:case1} shows snapshots of the swarm movement during its completion of the mission described by \eqref{eqn:case_spatel}.
It is seen that the swarm first satisfies $\varphi_3$, then $\varphi_2$ and then $\varphi_5$. 
\subsection{Case 2}
Now we set the initial condition to be uniformly distributed in the $NW$ quadrant of the $NE$ quadrant (see Fig. \ref{fig:case2} a).The MILP is solved in 43 seconds. The snapshots of the swarm movement are shown in Fig. \ref{fig:case2}. This time, the optimal plan is to satisfy $\varphi_2$ first, then $\varphi_3$ and finally $\varphi_6$.

\section{Conclusion and Future Work}
\label{sec:conclusions}
We presented a fully automated framework to synthesize controls that steer a fully actuated robotic swarm while  satisfying high-level spatio-temporal requirements. Such specifications are naturally expressed as spatial temporal logic formulae. A computationally efficient framework was introduced to determine the high level plan from which a low level  control strategy executes the swarm movements. 

Directions for future research include extending the framework to under actuated swarms and developing distributed control strategies for coordination of movements among robots. We also plan to incorporate machine learning methods from \cite{gol2014formal}  in order to synthesize control policies for more complex spatial patterns which are automatically learned from training data. Furthermore, we plan to create a graphical user interface in which a potential user can define required patterns for execution. The user would draw the patterns that they want to emerge and specify time requirements. The interface will use machine learning techniques to generate SpaTeL formulas for those patterns. These formulas will then be used by algorithms presented in this paper to synthesize control policies for the swarm.

\bibliography{references}

\begin{thebibliography}{10}
\providecommand{\url}[1]{#1}
\csname url@rmstyle\endcsname
\providecommand{\newblock}{\relax}
\providecommand{\bibinfo}[2]{#2}
\providecommand\BIBentrySTDinterwordspacing{\spaceskip=0pt\relax}
\providecommand\BIBentryALTinterwordstretchfactor{4}
\providecommand\BIBentryALTinterwordspacing{\spaceskip=\fontdimen2\font plus
\BIBentryALTinterwordstretchfactor\fontdimen3\font minus
  \fontdimen4\font\relax}
\providecommand\BIBforeignlanguage[2]{{%
\expandafter\ifx\csname l@#1\endcsname\relax
\typeout{** WARNING: IEEEtran.bst: No hyphenation pattern has been}%
\typeout{** loaded for the language `#1'. Using the pattern for}%
\typeout{** the default language instead.}%
\else
\language=\csname l@#1\endcsname
\fi
#2}}

\bibitem{mesbahi2010graph}
M.~Mesbahi and M.~Egerstedt, \emph{Graph theoretic methods in multiagent
  networks}.\hskip 1em plus 0.5em minus 0.4em\relax Princeton University Press,
  2010.

\bibitem{michael2008distributed}
N.~Michael, M.~M. Zavlanos, V.~Kumar, and G.~J. Pappas, ``{Distributed
  multi-robot task assignment and formation control},'' in \emph{Robotics and
  Automation, 2008. ICRA 2008. IEEE International Conference on}.\hskip 1em
  plus 0.5em minus 0.4em\relax IEEE, 2008, pp. 128--133.

\bibitem{cortes2002coverage}
J.~Cortes, S.~Martinez, T.~Karatas, and F.~Bullo, ``{Coverage control for
  mobile sensing networks},'' in \emph{Robotics and Automation, 2002.
  Proceedings. ICRA'02. IEEE International Conference on}, vol.~2.\hskip 1em
  plus 0.5em minus 0.4em\relax IEEE, 2002, pp. 1327--1332.

\bibitem{bullo2009distributed}
F.~Bullo, J.~Cort{\'e}s, and S.~Martinez, \emph{Distributed control of robotic
  networks: a mathematical approach to motion coordination algorithms}.\hskip
  1em plus 0.5em minus 0.4em\relax Princeton University Press, 2009.

\bibitem{kantor2006distributed}
G.~Kantor, S.~Singh, R.~Peterson, D.~Rus, A.~Das, V.~Kumar, G.~Pereira, and
  J.~Spletzer, ``{Distributed search and rescue with robot and sensor teams},''
  in \emph{Field and Service Robotics}.\hskip 1em plus 0.5em minus 0.4em\relax
  Springer, 2006, pp. 529--538.

\bibitem{thrun2005multi}
S.~Thrun and Y.~Liu, ``{Multi-robot SLAM with sparse extended information
  filers},'' in \emph{Robotics Research. The Eleventh International
  Symposium}.\hskip 1em plus 0.5em minus 0.4em\relax Springer, 2005, pp.
  254--266.

\bibitem{chen2005formation}
Y.~Chen and Z.~Wang, ``{Formation control: a review and a new consideration},''
  in \emph{Intelligent Robots and Systems, 2005.(IROS 2005). 2005 IEEE/RSJ
  International Conference on}.\hskip 1em plus 0.5em minus 0.4em\relax IEEE,
  2005, pp. 3181--3186.

\bibitem{Pimenta:2008aa}
L.~C. Pimenta, N.~Michael, R.~C. Mesquita, G.~A. Pereira, and V.~Kumar,
  ``Control of swarms based on hydrodynamic models,'' in \emph{Robotics and
  Automation, 2008. ICRA 2008. IEEE International Conference on}.\hskip 1em
  plus 0.5em minus 0.4em\relax IEEE, 2008, pp. 1948--1953.

\bibitem{egerstedt2001formation}
M.~Egerstedt and X.~Hu, ``Formation constrained multi-agent control,''
  \emph{IEEE TRANSACTIONS ON ROBOTICS AND AUTOMATION}, vol.~17, no.~6, p. 947,
  2001.

\bibitem{lee2014multi}
S.~G. Lee and M.~Egerstedt, ``Multi-robot control using time-varying density
  functions,'' \emph{arXiv preprint arXiv:1404.0338}, 2014.

\bibitem{yang2008multi}
P.~Yang, R.~Freeman, K.~M. Lynch, and Others, ``{Multi-agent coordination by
  decentralized estimation and control},'' \emph{Automatic Control, IEEE
  Transactions on}, vol.~53, no.~11, pp. 2480--2496, 2008.

\bibitem{kloetzer2007temporal}
M.~Kloetzer and C.~Belta, ``Temporal logic planning and control of robotic
  swarms by hierarchical abstractions,'' \emph{Robotics, IEEE Transactions on},
  vol.~23, no.~2, pp. 320--330, 2007.

\bibitem{gol2014formal}
E.~A. Gol, E.~Bartocci, and C.~Belta, ``{A formal methods approach to pattern
  synthesis in reaction diffusion systems},'' in \emph{Decision and Control
  (CDC), 2014 IEEE 53rd Annual Conference on}.\hskip 1em plus 0.5em minus
  0.4em\relax IEEE, 2014, pp. 108--113.

\bibitem{haghighi2015spatel}
I.~Haghighi, A.~Jones, Z.~Kong, E.~Bartocci, R.~Gros, and C.~Belta, ``{SpaTeL:
  a novel spatial-temporal logic and its applications to networked systems},''
  in \emph{Proceedings of the 18th International Conference on Hybrid Systems:
  Computation and Control}.\hskip 1em plus 0.5em minus 0.4em\relax ACM, 2015,
  pp. 189--198.

\bibitem{winfield2005formal}
A.~F. Winfield, J.~Sa, M.-C. Fern{\'a}ndez-Gago, C.~Dixon, and M.~Fisher, ``On
  formal specification of emergent behaviours in swarm robotic systems,''
  \emph{International journal of advanced robotic systems}, vol.~2, no.~4, pp.
  363--370, 2005.

\bibitem{chen2012formal}
Y.~Chen, X.~C. Ding, A.~Stefanescu, and C.~Belta, ``{Formal approach to the
  deployment of distributed robotic teams},'' \emph{Robotics, IEEE Transactions
  on}, vol.~28, no.~1, pp. 158--171, 2012.

\bibitem{tumova2014receding}
J.~Tumova and D.~V. Dimarogonas, ``{A receding horizon approach to multi-agent
  planning from local LTL specifications},'' in \emph{American Control
  Conference (ACC), 2014}.\hskip 1em plus 0.5em minus 0.4em\relax IEEE, 2014,
  pp. 1775--1780.

\bibitem{ulusoy2012robust}
A.~Ulusoy, S.~L. Smith, X.~C. Ding, and C.~Belta, ``{Robust multi-robot optimal
  path planning with temporal logic constraints},'' in \emph{Robotics and
  Automation (ICRA), 2012 IEEE International Conference on}.\hskip 1em plus
  0.5em minus 0.4em\relax IEEE, 2012, pp. 4693--4698.

\bibitem{diaz2015correct}
Y.~Diaz-Mercado, A.~Jones, C.~Belta, and M.~Egerstedt,
  ``Correct-by-construction control synthesis for multi-robot mixing,'' in
  \emph{Decision and Control (CDC), 2015 IEEE 54th Annual Conference on}.\hskip
  1em plus 0.5em minus 0.4em\relax IEEE, 2015, pp. 221--226.

\bibitem{maler_stl}
O.~Maler and D.~Nickovic, ``Monitoring temporal properties of continuous
  signals,'' in \emph{Formal Techniques, Modelling and Analysis of Timed and
  Fault-Tolerant Systems}.\hskip 1em plus 0.5em minus 0.4em\relax Springer,
  2004, pp. 152--166.

\bibitem{dokhanchi}
A.~Dokhanchi, B.~Hoxha, and G.~Fainekos, ``{On-line monitoring for temporal
  logic robustness},'' in \emph{Runtime Verification}.\hskip 1em plus 0.5em
  minus 0.4em\relax Springer, 2014, pp. 1--20.

\bibitem{raman}
V.~Raman, A.~Donz{\'e}, M.~Maasoumy, R.~M. Murray, A.~Sangiovanni-Vincentelli,
  and S.~A. Seshia, ``Model predictive control with signal temporal logic
  specifications,'' in \emph{Decision and Control (CDC), 2014 IEEE 53rd Annual
  Conference on}.\hskip 1em plus 0.5em minus 0.4em\relax IEEE, 2014, pp.
  81--87.

\bibitem{sadraddini2015robust}
S.~Sadraddini and C.~Belta, ``{Robust Temporal Logic Model Predictive
  Control},'' \emph{53rd Annual Conference on Communication, Control, and
  Computing (Allerton)}, 2015.

\bibitem{van2011reciprocal}
J.~Van Den~Berg, S.~J. Guy, M.~Lin, and D.~Manocha, ``Reciprocal n-body
  collision avoidance,'' in \emph{Robotics research}.\hskip 1em plus 0.5em
  minus 0.4em\relax Springer, 2011, pp. 3--19.

\end{thebibliography}

\end{document}